\documentclass[acmsmall,screen,nonacm]{acmart}
\newif\ifEnableExtend
\EnableExtendtrue

\newif\ifEnableExtendDONE
\EnableExtendDONEfalse

\usepackage[utf8]{inputenc}
\usepackage{url}
\usepackage{color}

\usepackage{amsmath}
\usepackage{amsthm}
\usepackage{xspace}
\usepackage{wasysym,pifont,marvosym}
\usepackage[colorinlistoftodos,prependcaption,textsize=small]{todonotes}
\usepackage{numprint}
\usepackage{csquotes}

\usepackage{relsize}
\usepackage[binary-units]{siunitx}
\usepackage{xcolor}

\usepackage{tikz}
\usetikzlibrary{calc}
\usepackage{pgfplots}
\usepgfplotslibrary{external}
\tikzexternaldisable
\newif\ifpdfplots
\pdfplotstrue

\newcommand{\ie}{i.e.}

\newcommand{\Queue}[1]{\Code{Q}}

\newcommand{\etal}{et~al.\xspace}

\usepackage{xargs}
\newcommandx{\uvc}[2][1=]{\todo[color=magenta!50,#1]{\sf \textbf{\"Umit:} #2}\xspace}

\newcommandx{\tobiasx}[2][1=]{\todo[color=black!50,#1]{\sf \textbf{Tobias:} #2}\xspace}
\newcommand{\lars}[2][]{\todo[color=purple!50,#1]{\sf \textbf{Lars:} #2}\xspace}

\newcommand{\ocut}{\ensuremath{\mathfrak{f}_c(\Partition)}}%
\newcommand{\ocon}{\ensuremath{\mathfrak{f}_\lambda(\Partition)}}%

\newcommand{\conset}{\ensuremath{\Lambda}}
\newcommand{\Partition}{\ensuremath{\mathrm{\Pi}}}%
\newcommand{\neighbors}{\ensuremath{\mathrm{\Gamma}}}%
\newcommand{\incnets}{\ensuremath{\mathrm{I}}}%
\newcommand{\pinsinpart}{\ensuremath{\mathrm{\Phi}}}
\newcommand{\adjblocks}{\ensuremath{\mathrm{B}}}
\newcommand{\con}{\ensuremath{\lambda}}

\setcopyright{acmcopyright}
\copyrightyear{2021}
\acmYear{2021}
\acmDOI{}
\acmPrice{15.00}
\acmISBN{xx}
\acmBooktitle{xx}
\title{More Recent Advances in (Hyper)Graph Partitioning}

\author{{\"{U}}mit V. {\c{C}}ataly{\"{u}}rek}
\email{umit@gatech.edu}
\affiliation{%
  \institution{Georgia Institute of Technology, Atlanta, GA}
  \streetaddress{North Avenue, Atlanta, GA 30332}
  \city{Atlanta}
  \state{Georgia}
  \country{United States of America}
}

\author{Karen D. Devine}
\email{devine.hpc@gmail.com}
\affiliation{%
  \institution{Sandia  National  Laboratories, ret.}
  \city{Albuquerque}
  \state{NM}
  \country{United States of America}
}

\author{Marcelo Fonseca Faraj}
\email{marcelofaraj@informatik.uni-heidelberg.de}
\affiliation{%
	\institution{Heidelberg University}
	\streetaddress{Im Neuenheimer Feld 205}
	\city{Heidelberg}
	\state{Baden-Württemberg}
	\country{Germany}
	\postcode{69120}
}

\author{Lars Gottesbüren}
\email{lars.gottesbueren@kit.edu}
\affiliation{%
  \institution{Karlsruhe Institute of Technology}
  \streetaddress{Am Fasanengarten 5}
  \city{Karlsruhe}
  \state{Baden-Württemberg}
  \country{Germany}
  \postcode{76131}
}

\author{Tobias Heuer}
\email{tobias.heuer@kit.edu}
\affiliation{%
  \institution{Karlsruhe Institute of Technology}
  \streetaddress{Am Fasanengarten 5}
  \city{Karlsruhe}
  \state{Baden-Württemberg}
  \country{Germany}
  \postcode{76131}
}

\author{Henning Meyerhenke}
\email{meyerhenke@hu-berlin.de}
\affiliation{%
  \institution{Humboldt-Universität zu Berlin}
  \streetaddress{Unter den Linden 6}
  \city{Berlin}
  \state{Berlin}
  \country{Germany}
  \postcode{10099}
}

\author{Peter Sanders}
\email{sanders@kit.edu}
\affiliation{%
  \institution{Karlsruhe Institute of Technology}
  \streetaddress{Am Fasanengarten 5}
  \city{Karlsruhe}
  \state{Baden-Württemberg}
  \country{Germany}
  \postcode{76131}
}

\author{Sebastian Schlag}
\email{sebastian_schlag@apple.com}
\affiliation{%
\institution{Apple Inc.}
 \city{Cupertino}
  \state{CA}
\country{United States of America}
}

\author{Christian Schulz}
\email{christian.schulz@informatik.uni-heidelberg.de}
\affiliation{%
  \institution{Heidelberg University}
  \streetaddress{Im Neuenheimer Feld 205}
  \city{Heidelberg}
  \state{Baden-Württemberg}
  \country{Germany}
  \postcode{69120}
}

\author{Daniel Seemaier}
\email{daniel.seemaier@kit.edu}
\affiliation{%
  \institution{Karlsruhe Institute of Technology}
  \streetaddress{Am Fasanengarten 5}
  \city{Karlsruhe}
  \state{Baden-Württemberg}
  \country{Germany}
  \postcode{76131}
}

\author{Dorothea Wagner}
\email{dorothea.wagner@kit.edu}
\affiliation{%
	\institution{Karlsruhe Institute of Technology}
	\streetaddress{Am Fasanengarten 5}
	\city{Karlsruhe}
	\state{Baden-Württemberg}
	\country{Germany}
	\postcode{76131}
}

\date{}
\renewcommand{\shortauthors}{{\c{C}}ataly{\"{u}}rek et al.}
\begin{document}

\begin{abstract}
In recent years, significant advances have been made in the design and evaluation of balanced (hyper)graph partitioning algorithms.
We survey trends of the last decade in practical algorithms for balanced (hyper)graph partitioning together with future research directions. Our work serves as an update to a previous survey on the topic \cite{SPPGPOverviewPaper}. In particular, the survey extends the previous survey by also covering hypergraph partitioning and streaming algorithms, and has an additional focus on parallel algorithms.
\end{abstract}
\vspace*{-1cm}
\maketitle


\tableofcontents
\vfill \pagebreak
\section{Introduction}

Graphs are a universal and widely used way to model relations between objects.
They are useful in a wide range of applications from social networks, simulation grids, road networks/route planning, (graph) neural networks, and many more.
With exploding data sets, the scalability of graph processing methods has become an increasing challenge.
As huge problem instances in the area become abundant, there is a need for scalable algorithms to perform analysis.
Often (hyper)graph partitioning is a key technology to make various algorithms scale in practice.
More precisely, scalable algorithms for various applications often require a subroutine that
 partitions a (hyper)graph into $k$ blocks of roughly equal size
such that the number of edges that run between blocks is minimized.
Balance is most often modeled using a \emph{balancing constraint} that demands that all block weights are below a given upper bound.
The number of applications that require such partitions of (hyper)graphs as key subroutine is huge.
Social network operators use graph
partitioning techniques to load balance their operation and make sure that site
response times are low \cite{socialhash}; efficient route planning algorithms
rely on a precomputation phase in which  a road network needs to be
partitioned~\cite{DellingGPW11}; scientific simulations that run on
supercomputers use (hyper)graph partitioning to balance load and minimize
communication between
processors~\cite{DBLP:conf/europar/FietzKSSH12,DBLP:conf/cluster/SchloegelKK01,catalyurek2009repartitioning};
partitioning (graph) neural networks is helpful to speedup training times
\cite{DBLP:journals/corr/abs-2112-15345}; and (hyper)graph partitioning helps
to improve chip placements in VLSI design \cite{alpert1995rdn,Papa2007}.
On the other side, the problem is NP-hard \cite{Garey1974} even for unweighted trees of maximum degree three \cite{an2020complexity} and no constant-factor approximation algorithms exist~\cite{andreev2006balanced}.
Thus, heuristic algorithms are used in practice.
The purpose of this paper is to give a structured overview of the rich literature, with a clear emphasis on
explaining key ideas and discussing work that has been published in the last decade and is missing in other overviews.
More precisely, our work serves as an update to a previous generic survey on the topic \cite{SPPGPOverviewPaper}; in particular we also extend the scope of the survey to hypergraph partitioning algorithms.

There have been other (older) surveys on the topic.
The book by Bichot and Siarry~\cite{GPOverviewBook} covers
techniques for graph partitioning such as the multilevel method, metaheuristics, parallel methods, and hypergraph partitioning, as well as applications of graph partitioning.
The survey by Schloegel \etal~\cite{SchloegelKarypisKumar03graph} was published around the turn of the millennium and has a focus on
techniques for scientific computing, including algorithms for adaptive and dynamic simulations and process mapping algorithms. Static algorithms as well as formulations with multiple objectives and constraints are also discussed.   Monien \etal~\cite{MonienPS07approximation} discuss multilevel algorithms. Their
description has a focus on matching-based coarsening and local search that use vertex-swapping heuristics. Kim \etal \cite{Kim11} cover memetic algorithms.
Abbas~\etal~\cite{DBLP:journals/pvldb/AbbasKCV18} describe and categorize
streaming graph partitioning techniques based on their assumptions, objectives
and costs, and perform an experimental comparison of the
algorithms for different applications and datasets.
The last generic survey on the topic by
Bulu{\c{c}}~\etal~\cite{SPPGPOverviewPaper} covers a wide range of techniques
and practical algorithms that have been published in or before 2013.
For hypergraph partitioning there exist two older surveys \cite{alpert1995rdn,Papa2007}. Alpert and Kahng~\cite{alpert1995rdn} discuss min-cut and ratio cut bipartitioning formulations along with multi-way extensions, constraint-driven partitioning and partitioning with replication. Papa and Markov~\cite{Papa2007} discuss practical applications of hypergraph partitioning, exact algorithms, and various local search heuristics for the problem as well as software packages and benchmarks. Schlag~\cite{DBLP:phd/dnb/Schlag20} presents a recent overview of hypergraph techniques in his dissertation.

Our survey is structured as follows. We start by introducing the problems and other preliminaries in Section~\ref{sec:prelim}. Then in Section~\ref{sec:applications}, we focus on new applications that
emerged in the last decade. We continue with novel sequential techniques in Section~\ref{sec:sequentialtechniques}.
Then we continue with covering parallel algorithms in Section~\ref{sec:parallelalgos}.
We describe experimental methodology in Section~\ref{sec:experimentalmethodology}.
We conclude with future challenges in Section~\ref{sec:futchallenges}.

\section{Overview/Classification}
The field is currently very active. On the one side there are \emph{streaming algorithms} that are very fast and consume little memory yet yield only low quality solutions.
These algorithms assume that internal memory of a single machine proportional to the number of vertices is available and typically load nodes and their neighborhood one by one to directly make an assignment to a block. Hence, the edges of a graph do not have to fit into the memory of the machine.
On the other side, there is a wide range of sequential internal-memory algorithms that tackle partitioning problems that fit into the main memory of a single machine without using parallelism.
Recent advances in this area provide a rich set of algorithms that are able to compute partitions of very high quality.
These high-quality algorithms often also have shared-memory parallel counterparts.
While the sequential internal memory/parallel shared-memory non-metaheuristic algorithms in this area use a reasonable amount of time for most applications,
there are also memetic or evolutionary algorithms that invest a lot of resources to achieve even higher quality. Note that memetic or evolutionary algorithms can be sequential internal-memory, shared-memory or even distributed memory parallel.
Due to their high running time, such algorithms are typically used only on graphs having a few hundred thousand nodes.
Exact solvers are currently able to solve only very small instances while also requiring %
a lot of time
to solve an instance even for few %
blocks. For larger numbers of blocks, the exact solvers currently do not work well.
Distributed parallel algorithms scale well to large instances, but if these tools do not implement multilevel strategies, they typically have much lower quality than sequential internal memory partitioning algorithms.
Even when such algorithms use the multilevel scheme, solution quality of the algorithms operating in this model of computation is typically worse compared to their internal memory/shared-memory counterparts.
However, this mode of computation has the advantage that huge instances can be partitioned very quickly and also that it enables researchers to partition huge graphs on cheap machines.
Lastly, researchers also work on algorithms for other platforms like GPUs to be able to fully use the capabilities of existing hardware.
However, GPUs have limited memory size and current algorithms running on a GPU compute partitions that cut significantly more edges than high-quality internal~memory~schemes.

\section{Preliminaries}
\label{sec:prelim}

\subsection{Notation}\label{sec:notation}
\paragraph{Hypergraphs \& Graphs}
A \textit{weighted undirected hypergraph} $H=(V,E,c,\omega)$ is defined as a set of $n$ vertices $V$ and a
set of $m$ hyperedges/nets $E$ with vertex weights $c:V \rightarrow \mathbb{R}_{>0}$ and net
weights $\omega:E \rightarrow \mathbb{R}_{>0}$, where each net $e$ is a subset of the vertex set $V$ (i.e., $e \subseteq V$).
The vertices of a net are called \emph{pins}.
We extend $c$ and $\omega$ to sets in the natural way, i.e., $c(U) :=\sum_{v\in U} c(v)$ and $\omega(F) :=\sum_{e \in F} \omega(e)$.
A vertex $v$ is \textit{incident} to a net $e$ if $v \in e$. $\mathrm{I}(v)$ denotes the set of all incident nets of $v$.
The set $\neighbors(v) := \{ u~|~\exists e \in E : \{v,u\} \subseteq e\}$ denotes the neighbors of $v$.
The \textit{degree} of a vertex~$v$ is $d(v) := |\mathrm{I}(v)|$.
 We assume hyperedges to be sets rather than multisets, i.e., a vertex can only be contained in a hyperedge \emph{once}.
Nets of size one are called \emph{single-vertex} nets.
Given a subset $V' \subset V$, the \emph{subhypergraph} $H_{V'}$ is defined as $H_{V'}:=(V', \{e \cap V'~|~e \in E : e \cap V' \neq \emptyset \})$.

A \textit{weighted undirected graph} $G=(V,E,c,\omega)$ is defined as a set of $n$ vertices $V$ and a
set of $m$ and edges $E$ with vertex weights $c:V \rightarrow \mathbb{R}_{>0}$ and edge
weights $\omega:E \rightarrow \mathbb{R}_{>0}$.
In contrast to hypergraphs, the size of the edges is restricted to two. Let $G=(V,E,c,\omega)$ be a weighted (directed) graph. We use \emph{hyperedges/nets} when referring to hypergraphs and \emph{edges} when referring to graphs.
However, we use the same notation to refer to vertex weights $c$,
edge weights $\omega$, vertex degrees $d(v)$, and the set of neighbors $\neighbors$.
In an undirected graph, an edge $(u,v) \in E$ implies an edge $(v,u) \in E$ and $\omega(u,v) = \omega(v,u)$.

\paragraph{Partitions and Clusterings}
A \emph{$k$-way partition} of a (hyper)graph $H$ is a partition of its vertex set into $k$ \emph{blocks} $\Partition = \{V_1, \dots, V_k\}$
such that $\bigcup_{i=1}^k V_i = V$, $V_i \neq \emptyset $ for $1 \leq i \leq k$, and $V_i \cap V_j = \emptyset$ for $i \neq j$.
We call a $k$-way partition $\Partition$ \emph{$\varepsilon$-balanced} if each block $V_i \in \Partition$ satisfies the \emph{balance constraint}:
$c(V_i) \leq L_{\max} := (1+\varepsilon)\left\lceil \frac{c(V)}{k} \right\rceil$ for some parameter $\mathrm{\varepsilon}$.%
\footnote{The $\lceil\cdot\rceil$ in this definition ensures that there is always a feasible solution for inputs with unit vertex weights. For general weighted inputs, there is no commonly accepted way how to deal with feasibility; see also \cite{HMS21,kaminpar}.}
We call a block $V_i$ \emph{overloaded} if $c(V_i) > L_{\max}$.
For each net $e$, $\conset(e) := \{V_i~|~ V_i \cap e \neq \emptyset\}$ denotes the \emph{connectivity set} of $e$.
The \emph{connectivity} $\con(e)$ of a net $e$ is the cardinality of its connectivity set, i.e.,  $\con(e) := |\conset(e)|$.
A net is called a \emph{cut net} if $\con(e) > 1$; otherwise  (i.e., if $|\mathrm{\lambda}(e)|=1$ ) it is called an \emph{internal} net.
A vertex $u$ that is incident to at least one cut net is called a  \emph{border vertex}.
The number of pins of a net $e$ in block $V_i$ is defined as  $\pinsinpart(e,V_i) := |\{V_i \cap e \}|$.
A block $V_i$ is \emph{adjacent} to a vertex $v \notin V_i$ if $\exists e \in  \incnets(v) : V_i \in \conset(e)$.
We use $\adjblocks(v)$ to denote the set of all blocks adjacent to $v$.
Given a $k$-way partition $\Partition$ of $H$, the \emph{quotient graph}
$Q := (\Partition, \{(V_i,V_j)~|~\exists e \in E : \{V_i,V_j\} \subseteq  \conset(e)\})$ contains an edge between each pair of adjacent blocks.
A \emph{clustering} $C = \{C_1, \dots, C_l\}$ of a hypergraph is a partition of its vertex set. In contrast to a $k$-way partition,
the number of clusters is not given in advance, and there is no balance constraint on the actual sizes of the clusters $C_i$.

A \emph{$k$-way (hyper)edge partition} of a (hyper)graph $H$ is a partition of its (hyper)edge set into $k$ \emph{blocks} $\Partition = \{E_1, \dots, E_k\}$
such that $\bigcup_{i=1}^k E_i = E$, $E_i \neq \emptyset $ for $1 \leq i \leq k$, and $E_i \cap E_j = \emptyset$ for $i \neq j$.
We call a $k$-way (hyper)edge partition $\Partition$ \emph{$\varepsilon$-balanced} if each block $E_i \in \Partition$ satisfies the \emph{balance constraint}:
$\omega(E_i) \leq (1+\varepsilon)\lceil \frac{\omega(E)}{k} \rceil$ for some parameter $\mathrm{\varepsilon}$. %
A vertex is called a \emph{cut vertex} if it contains two or more edges in different blocks.
The (weighted) \emph{vertex-cut} of a \emph{$k$-way (hyper)edge partition} is the total weight of cut vertices.

\subsection{The \texorpdfstring{$k$}{k}-way  (Hyper)Graph Partitioning Problem}\label{sec:hgp}
\begin{figure}
\includegraphics[width=150pt]{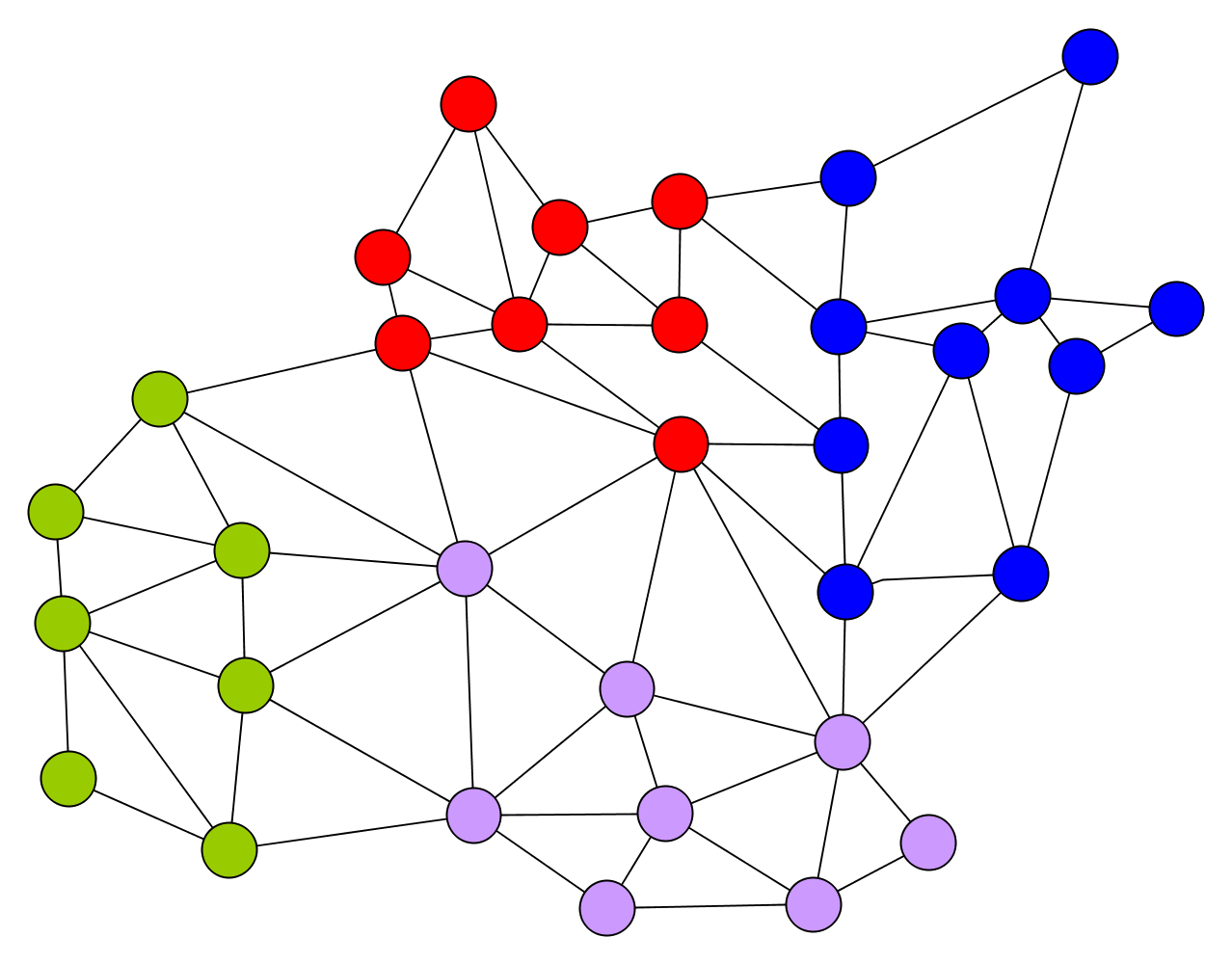}\hspace*{1cm}
\includegraphics[width=150pt]{img/hgp}
\caption{Left: An example graph that is partitioned into four blocks indicated by the colors. The partition has an edge cut of 17, the block weights $|V_i|$ are green (7), red (8), blue (10), purple (8). Right: An example hypergraph that is partitioned into four blocks indicated by the colors. Hyperedges are indicated by black lines or colored blobs around the vertices (if more than two nodes are contained in an hyperedge). Blocks have the same weight as in the graph case. The connectivity objective is 22 and the  number of hyperedges cut is 20. }
\label{fig:gp}
\end{figure}

\paragraph{Problem Definition}
 The \emph{$k$-way (hyper)graph partitioning problem} is to find an $\varepsilon$-balanced $k$-way partition $\mathrm{\Pi}$ of a (hyper)graph $H=(V,E,c,\omega)$ that
 \emph{minimizes} an objective function over the cut nets for some value of~$\varepsilon$.
The two most commonly used cost functions in case we are dealing with hypergraphs are the \emph{cut-net} metric $\ocut := \sum_{e \in E'} \omega(e)$ and the
\emph{connectivity} metric $\ocon := \sum_{e\in E'} (\lambda(e) -1)~\omega(e)$, where $E'$ is the \emph{cut-set} (i.e., the set of all cut nets)~\cite{donath1988logic,DBLP:journals/jpdc/DeveciKUC15}.
While the cut-net metric sums the weights of all nets that connect more than one block of the partition $\Partition$, the connectivity
metric additionally takes into account the actual number $\con$ of blocks connected by the cut nets.
When partitioning graphs, the objective is often to minimize $\sum_{i < j} \omega(E_{ij})$ (weight of all cut edges), where
  $E_{ij} \coloneqq \{\{u, v\} \in E \mid u \in V_i, v \in V_j \}$.
Note that for graphs the objective functions $\ocut$ and $\ocon$ revert to edge-cut (i.e., the sum of the weights of those edges that have endpoints in different blocks). Apart from these common
cost functions, other specialized objective functions exist \cite{SPPGPOverviewPaper}.
Figure~\ref{fig:gp} shows an example partition of a graph.
Throughout the paper, we use the term high-quality if the respective objective function is small compared to other tools, and low-quality if the opposite is the case.

\section{Applications}
\label{sec:applications}

Research on graph partitioning algorithms has always been motivated by its numerous applications.
Some traditional applications are parallel processing (e.g., in scientific computing), VLSI design, route planning and image segmentation; see also the previous survey~\cite{SPPGPOverviewPaper}.
Yet, in recent years several new applications have emerged.
\paragraph{Distributed Databases and Query Optimization}

A prominent application is distributed database sharding~\cite{Application:DistributedDB, DBLP:journals/vldb/KumarQDK14, socialhash, DBLP:journals/fgcs/YangWCC18, atrey2020unifydr}.
Vertices represent data records, hyperedges represent queries that access records, and blocks correspond to shards (machines).
Minimizing connectivity thus corresponds to minimizing the average number of shards involved in a query, which optimizes the latency and query processing time.

A similar application is to boost IO throughput in the Google search engine backend by improving cache utilization.
Archer \etal~\cite{google-search} initialize a voting table that assigns search requests, using graph partitioning on a bipartite graph where vertices are search terms and queries, and an edge exists for each term contained in a query.
Subsequent simulation-and-refinement further boosts the prediction accuracy (item in cache) of the voting table.

Li \etal~\cite{li2021les3} use graph partitioning to construct a set similarity search index.
The approach first filters candidate sets using the index, then checks the remaining candidates brute-force.
Vertices represent sets, edges connect sets that are $k$-nearest neighbors (or within a similarity threshold, depending on the query type), and cut size is correlated with pruning efficiency.

\paragraph{Data Distribution and Scheduling}

Djidjev \etal~\cite{DBLP:journals/algorithms/DjidjevHMNN19} use graph partitioning to accelerate and parallelize the computation of density matrices for molecular dynamics simulations.
Lattice Boltzmann Fluid Flow simulations~\cite{heuvelinecoop, DBLP:conf/sc/GodenschwagerSBKR13, DBLP:journals/cma/KrauseKAKDKGHMT21} fall into the classic load balancing and communication minimization application category of graph partitioning.
Yu \etal~\cite{yu2020numa} consider sparse matrix vector multiplication~\cite{PaToH}, however focused on shared-memory machines with many NUMA nodes instead of distributed systems.

Parallel graph processing systems such as Pregel~\cite{mccune-tlav-2015} or Giraph~\cite{giraph} have become wide-spread tools for network analysis tasks.
These systems employ a think-like-a-vertex or think-like-an-edge programming paradigm, requiring a balanced partition of vertices or edges across machines.
Streaming approaches~\cite{hoang2019cusp, DBLP:conf/icde/MartellaLLS17} based on label propagation result in better performance than traditional range-based or hash-based partitioning.

\paragraph{Quantum Circuit Simulation}

Several papers employ nested dissection on hypergraphs to find good contraction trees, in order to speed up the simulation of quantum circuits on classical non-quantum machines~\cite{huang2020classical, gray2021hyper, pan2021simulating}.
This can be used to experimentally verify the correctness of a quantum circuit, and push back premature claims of having achieved quantum supremacy.

\paragraph{SAT Solving}

Boolean satisfiability (SAT) formulas can be represented as hypergraphs.
One representation is the dual representation with clauses as vertices and variables as hyperedges, consisting of clauses that contain the variable (either negated or not).
Mann and Papp~\cite{DBLP:journals/ijait/MannP17} use balanced bipartitioning to identify variables that branching solvers should focus on assigning first, such that the formula is split into two once variables in the cut are assigned.
The two sub-formulas can be solved independently, which is expected to be faster.
Most SAT solvers employ a heuristic (VSIDS) that assigns priorities (frequency in observed conflict clauses) to variables, selecting the highest priority to branch on next.
The authors consider two variants, the latter of which is more successful: force split the formula first and use VSIDS for tie-breaking, or initialize the priorities with the partition and let VSIDS overrule the decision.

\paragraph{Miscellaneous}

Lamm \etal~\cite{DBLP:conf/wea/LammS015} formulate recombine-operators in an evolutionary framework for independent sets based on separator decompositions and graph partitions.
Given two independent sets, their vertices in the two blocks of a separator decomposition are exchanged, yielding two offspring independent sets that are refined using local search.

Kumar \etal~\cite{DBLP:conf/accv/KumarCT14} use a clustering formulation to compute trajectories of moving objects in videos.
Detected objects in a frame correspond to vertices and each block corresponds to an object.
The authors enforce special constraints such that each partition contains only one object per frame and such that objects do not jump between frames.

Yao \etal~\cite{DBLP:conf/icc/YaoHZLN15} consider the placement of control units in software-defined networks to minimize the average latency of messages from switches (nodes) to controllers.
First the number of necessary controllers is estimated based on message volume capacities, then the switch graph is partitioned and one control unit is placed in each block, using a separate routine to perform placement inside the blocks.

Quantum chemical simulations such as \emph{density functional theory} exhibit quadratic time complexity, which is why calculations on largely independent sub-systems are used to approximate and accelerate the process.
Von Looz \etal~\cite{DBLP:conf/wea/LoozWJM16} propose to use graph partitioning to automatize the sub-system construction process, where small edge cuts correspond to small introduced calculation~errors.

\section{Sequential Techniques}
\label{sec:sequentialtechniques}

\subsection{Classic (Hyper)Graph Partitioning Techniques}

This section discusses partitioning techniques repeatedly used in this survey. We briefly outline
the multilevel paradigm and the most common local search algorithms. These algorithms move
vertices according to gain value. The gain value $g_u(V_j)$ reflects the change in the objective function when we move a vertex $u$
from its current to a target block $V_j$ (e.g., reduction in the edge cut).

\paragraph{The Multilevel Paradigm.}
The most successful approach to solve the (hyper)graph partitioning problem is the \emph{multilevel} paradigm. It consists of
three phases. In the \emph{coarsening} phase, a hierarchy of successively smaller and structurally similar (hyper)graphs are created
by contracting matchings or clusters of vertices. Once the (hyper)graph is small enough, an \emph{initial partitioning} algorithm
obtains a partition of the coarsest (hyper)graph. In the \emph{uncoarsening} phase, the partition is projected to the next larger (hyper)graph in the hierarchy,
and, at each level, \emph{local search} algorithms improve the objective function (e.g., edge cut).
Figure~\ref{fig:multilevel} illustrates the multilevel paradigm.

A contraction of several vertices into a supervertex aggregates their weight in the supervertex.
To further reduce the size of the coarser (hyper)graph, one can also remove all edges that become identical
except for one, at which their weight is aggregated (self-loops are discarded).
This way, one obtains a partition with the same balance and cut properties when projecting a partition to the next larger hypergraph in the hierarchy.

\begin{figure}[t!]
\includegraphics[width=9cm]{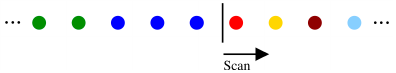}
        \begin{tabular}{ccc}
                \includegraphics[width=3cm]{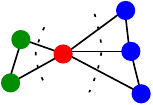} & \begin{minipage}{.025\textwidth}\vspace*{-2cm}\textbf{$\rightarrow$}\end{minipage} &
                \includegraphics[width=3cm]{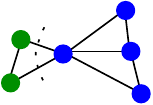}
        \end{tabular}

\caption{Label propagation algorithm in graphs. Top: nodes are scanned in a specific order. Colors of nodes indicate their block.  The algorithm initially assigns each node its own label and then visits the vertices in some order, assigning the current vertex the label that is most frequent in its neighborhood. }
\label{fig:labelpropagation}
\end{figure}

\paragraph{The Label Propagation Algorithm.}
The label propagation algorithm, illustrated in Fig.~\ref{fig:labelpropagation}, was originally proposed to detect community structures in large-scale networks~\cite{labelpropagationclustering,LABEL-PROPAGATION-ML}
but was also used as a refinement technique in the partitioning context~\cite{hMetis-K}.
The label propagation algorithm works in rounds, and each vertex $u$ is associated with a label $L[u]$.
Initially, each vertex is assigned its own label (i.e., $L[u] = u$).
In each round, the vertices are visited in some order, and whenever a vertex $u$ is visited, it adapts its
label to the label appearing most frequently in its neighborhood (ties are broken randomly).
The algorithm proceeds until it reaches a predefined number of rounds or none of the vertices changes its label
in a round.
The algorithm can be used as a clustering algorithm in the coarsening phase~\cite{DBLP:conf/wea/MeyerhenkeSS14}
or a local search algorithm when the current $k$-way partition is used for the initial label assignment.

\paragraph{The Fiduccia-Mattheyses Algorithm.}
Fiduccia and Mattheyses~\cite{fiduccia1982lth} presented the first linear-time heuristic for the
balanced bipartitioning problem (referred to as FM algorithm).
The algorithm works in passes, and a vertex can change its block at most once in a pass.
The FM algorithm uses two priority queues (one for each block). Initially, it inserts each boundary
vertex into the priority queue, along with the gain of moving the vertex to the opposite block.
Each step repeatedly performs the move with the highest gain that does not violate the balance constraint
and, subsequently, updates the gain of all non-moved neighbors.
A pass ends when all vertices are moved, or the balance constraint prevents further moves.
The FM algorithm also performs negative gain moves and is therefore able to escape from local optima.
In the end, it reverts to best seen solution during the~pass.

\ifEnableExtendDONE
\noindent 2016 $k$-way Hypergraph Partitioning via $n$-Level Recursive Bisection \cite{DBLP:conf/alenex/SchlagHHMS016}\\
2020 High-Quality Hypergraph Partitioning \cite{DBLP:phd/dnb/Schlag20} \\
2021 Deep Multilevel Graph Partitioning \cite{DBLP:journals/corr/abs-2105-02022} \\
\fi

\subsection{Recent Advances in Multilevel Partitioning}

This section discusses recent developments in the multilevel partitioning context.
We start with two new multilevel partitioning schemes preferable in situations where either
the number of blocks is large or high solution quality is required.
We then take a closer look at the different phases of multilevel scheme and
highlight recent algorithmic~improvements.

\paragraph{Deep Multilevel Partitioning.}
A $k$-way partition of a (hyper)graph can be obtained either by \emph{recursive bipartitioning} (RB) or \emph{direct $k$-way partitioning}. The former computes
a bipartition of the input (hyper)graph and then recurses on both blocks until the (hyper)graph is divided into the desired number of blocks. The latter
partitions the (hyper)graph directly into $k$ blocks and applies $k$-way local search algorithms to improve the solution.

Recently, these approaches have been generalized to \emph{deep multilevel partitioning}~\cite{kaminpar}.
The approach continues coarsening until only $2X$ vertices are left (where $X$ is an input parameter)
and computes an initial bipartition of the coarsest (hyper)graph. In the uncoarsening phase, it bipartitions a block when its size becomes larger than $2X$
vertices as long as there are less than~$k$~blocks.

\begin{figure}[t!]
  \centering
  \includegraphics[width=0.9\textwidth]{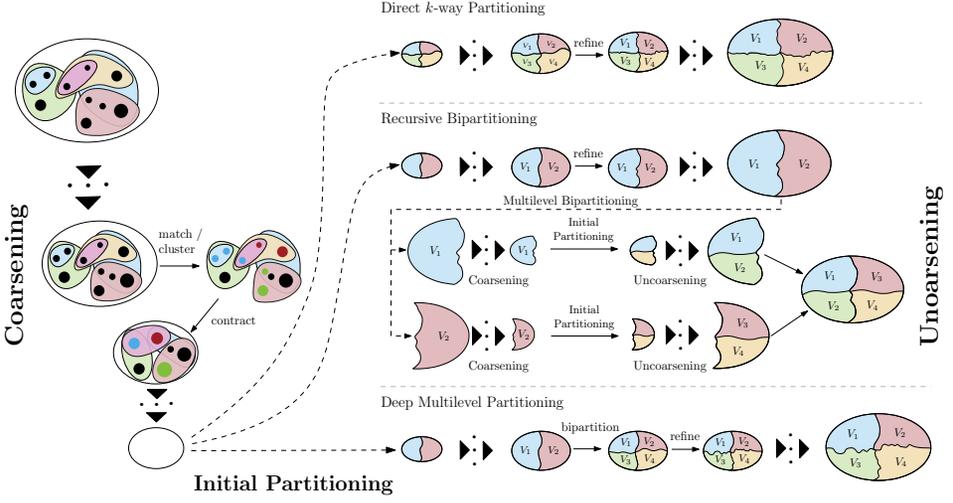}
  \caption{The multilevel paradigm and its different instantiations to obtain a $k$-way partition.}
  \label{fig:multilevel}
  \vspace{-0.25cm}
\end{figure}

The deep multilevel approach combines the strengths of the RB and direct $k$-way scheme: It recursively bipartitions
blocks of size $O(1)$ in the uncoarsening phase and thereby enables using $k$-way local search algorithms, while the
RB scheme recursively bipartitions blocks of size $O(k)$ and direct $k$-way reverts to RB for initial
partitioning~\cite{kPaToH, hMetis-K, KaHyPar-K, dissSchulz}.
Figure~\ref{fig:multilevel} illustrates the different instantiations
of the multilevel scheme.
\paragraph{$n$-Level (Hyper)Graph Partitioning.}
The depth of the multilevel hierarchy offers a trade-off between running time and solution quality of a multilevel algorithm~\cite{MLPart}.
More levels provide \textcquote{MLPart}{more opportunities to refine the current solution} at different granularities but require fast local
search algorithms to achieve reasonable running times.
The most extreme version of the multilevel paradigm is to contract only a single vertex at each level, which induces a hierarchy with
almost $n$ levels.
This technique was first studied by Osipov~\etal~\cite{kaspar} for graph partitioning and later improved further
by Schlag~\etal~\cite{KaHyPar-K, DBLP:conf/alenex/SchlagHHMS016, DBLP:phd/dnb/Schlag20} for hypergraph partitioning.
The approach is made feasible by using a
highly-localized variant of the classical FM algorithm~\cite{fiduccia1982lth} that initializes the priority queue only with
the uncontracted nodes and expands the search to neighbors of moved nodes~\cite{kaffpatr,kaspar}. Furthermore, the implementation uses a gain cache to avoid expensive recomputations of
move gains~\cite{KaHyPar-K} (reducing the running time of the FM algorithm by 45\%) and an adaptive stopping rule that terminates a search early if it becomes unlikely to find further
improvements~\cite{kaspar} (reducing the running time by an order of magnitude).
A parallel version of the n-level scheme exists~\cite{n-level-mt-kahypar-tr,MT-KAHYPAR-Q} that uncontracts a fixed number of vertices in parallel on each level (instead of a single vertex)..

\paragraph{Coarsening}
The coarsening phase aims to compute successively coarser approximations of the input (hyper)graph
such that its structural properties are maintained and act in some sense as a filter that removes as much unnecessary information from
the search space as possible~\cite{Walshaw2003}. In the past years, research focused on clustering techniques for complex
networks and enhancing the coarsening process with information about the community structure of the (hyper)graph.

Meyerhenke~\etal~\cite{DBLP:journals/heuristics/MeyerhenkeS016} use the size-constraint label propagation algorithm~\cite{labelpropagationclustering} to compute a vertex clustering,
which is then contracted to form the multilevel hierarchy. Clustering algorithms can reduce the size of complex networks (power-law degree distribution) more efficiently than
previously used matching-based approaches. The latter has the problem that vertices incident to high-degree vertices often remain unmatched. Therefore,
Davis~\etal~\cite{davis2020algorithm} propose several techniques to reduce the number of unmatched vertices. Two non-adjacent unmatched vertices are matched
if they share a common neighbor (also known as $2$-hop matching~\cite{karypis1998fast}) or if they are adjacent to two vertices that are already matched.
Additionally, one can extend two matched vertices with an unmatched neighbor ($3$-way match).

The following two techniques recompute the (hyper)edge weights of the input (hyper)graph and use them as an input for a multilevel algorithm.
The new weights encode more information about the importance of a (hyper)edge and should prevent a coarsening algorithm from collapsing
(hyper)edges into a single vertex that are in the cut set of a good partition.
Glantz~\etal~\cite{DBLP:journals/jea/GlantzMS16}
define the weight of an edge $e \in E$ as the minimum conductance value of all
bipartitions induced by a spanning tree in which $e$ is a cut edge.
The conductance of a bipartition $(V_1, V_2)$ is
$cond(V_1,V_2) := \frac{\omega(E'(V_1,V_2))}{\min \{ vol(V_1), vol(V_2) \}}$ where $vol(V_i)$ is the weight of all edges incident to vertices in $V_i$.
Chen and Safro~\cite{DBLP:journals/mmas/ShaydulinCS19} use the maximum algebraic distance~\cite{chen-safro-algdist-full, safro:relaxml}
between two pins of a hyperedge as~its~weight.

Lotfifar and Johnson~\cite{DBLP:conf/europar/LotfifarJ15} compute a hyperedge clustering and assign a vertex to the cluster containing
most of its incident nets. Their matching-based coarsening algorithm uses the restriction that matched vertices must be part of the same cluster.
Heuer and Schlag~\cite{hs2017sea} also use vertex clustering to restrict contractions to densely-coupled regions of a hypergraph.
Their algorithm transforms the hypergraph into its bipartite graph representation and then uses the Louvain algorithm~\cite{Louvain} maximizing the modularity objective function
to exploit its community structure.

Shaydulin and Safro~\cite{DBLP:conf/wea/ShaydulinS18} use ideas from algebraic multigrid as well as stable matching in which the preference lists determine a good combination of nodes for aggregation-based coarsening. 
The authors integrate their approaches into the Zoltan tool. 
Their experimental results with Zoltan, hMetis and PaToH demonstrate that given the same refinement, the proposed schemes are at
least as effective as traditional matching-based schemes, while outperforming them on many
instances.

\ifEnableExtendDONE
2015  A Multi-level Hypergraph Partitioning Algorithm Using Rough Set Clustering \cite{DBLP:conf/europar/LotfifarJ15} \\
2016 (conf 2013) Partitioning (hierarchically clustered) complex networks via size-constrained graph clustering\cite{DBLP:journals/heuristics/MeyerhenkeS016} \\
2016 (conf 2014) Tree-Based Coarsening and Partitioning of Complex Networks\cite{DBLP:journals/jea/GlantzMS16} \\
2017  Improving Coarsening Schemes for Hypergraph Partitioning by Exploiting Community Structure \cite{hs2017sea} \\
2019  Relaxation-Based Coarsening for Multilevel Hypergraph Partitioning \cite{DBLP:journals/mmas/ShaydulinCS19} \\
2020 Algorithm 1003: Mongoose, a graph coarsening and partitioning library \cite{davis2020algorithm} \\
2020 Balanced Coarsening for Multilevel Hypergraph Partitioning via Wasserstein Discrepancy \cite{guo2021balanced}
\fi

\paragraph{Initial Partitioning}
Coarsening usually proceeds until $\Omega(k)$ vertices remain.
Partitioners based on the direct $k$-way scheme
often use multilevel recursive bipartitioning to obtain an initial $k$-way partition of the coarsest (hyper)graph~\cite{kPaToH, hMetis-K, KaHyPar-K, dissSchulz}.
Heuer~\cite{KAHYPAR-IP} showed that this leads to better initial partitions than flat partitioning techniques.
To obtain an initial bipartition, many partitioners run a \emph{portfolio} of different flat bipartitioning algorithms multiple times
(e.g., greedy graph growing, random or spectral partitioning) followed by label propagation or FM local search and
continue uncoarsening with the best bipartition out of these runs~\cite{KaHyPar-K, karypis1998fast, PaToHManual}.
Preen and Smith~\cite{DBLP:journals/tec/PreenS19} showed that the decision
when to stop coarsening is often instance-dependent and proposed an adaptive stopping criterion (instead of using the same constant number of vertices for all instances).

\ifEnableExtendDONE
2016 $k$-way Hypergraph Partitioning via $n$-Level Recursive Bisection \cite{DBLP:conf/alenex/SchlagHHMS016} \\
2019 Evolutionary $n$-level Hypergraph Partitioning with Adaptive Coarsening \cite{DBLP:journals/tec/PreenS19}
\fi

\paragraph{Refinement.}
\label{lab:flowbasedrefinment}

\begin{figure}[t]
\begin{minipage}[t]{4.75cm}
\includegraphics[width=4.45cm]{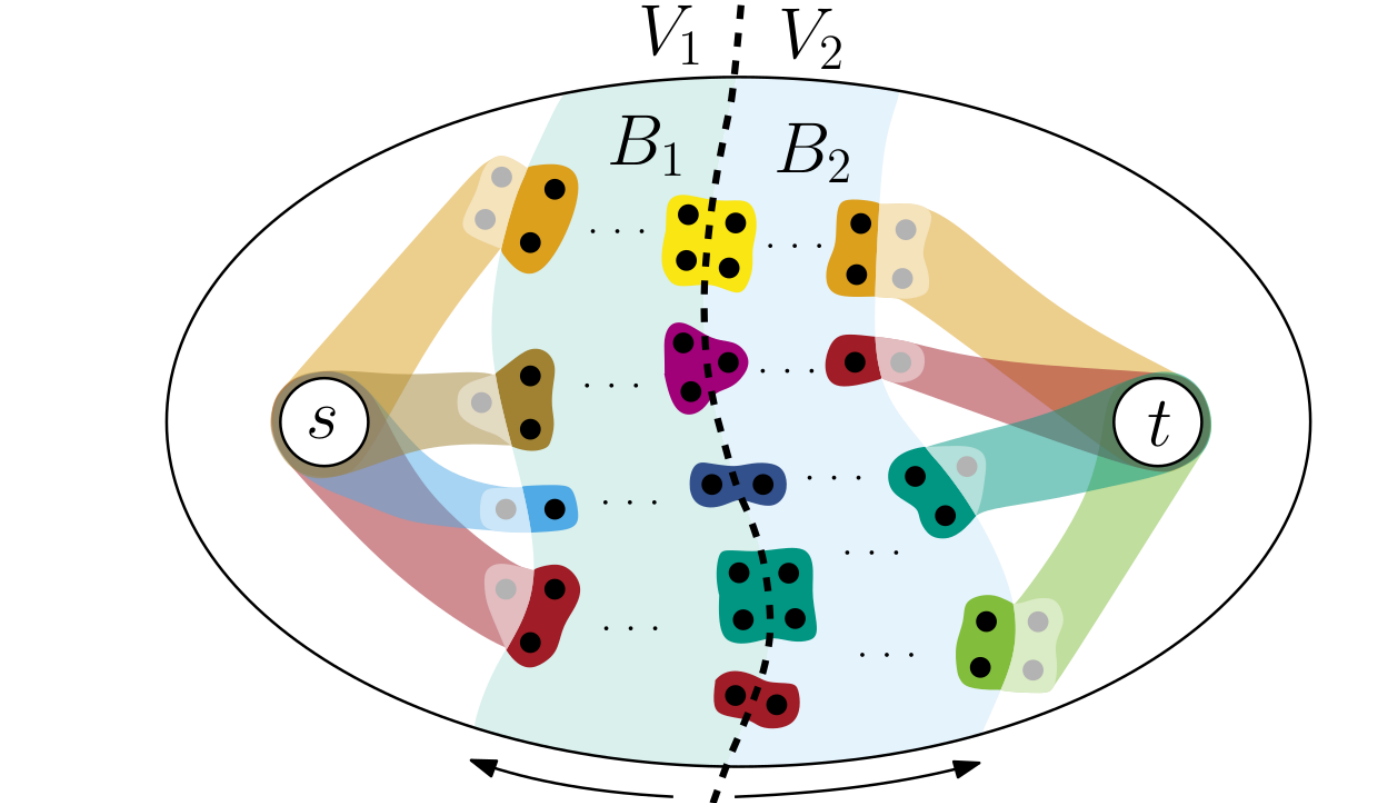}
\end{minipage}
\hspace*{.5cm}
\begin{minipage}[t]{8.5cm}
\includegraphics[width=8.35cm]{img/fbb_algorithm.pdf}
\end{minipage}
\caption{Left: a flow network that is constructed around a bipartition of a hypergraph. Running a max-flow min-cut algorithm on this network is used to find a reduced cut in the original network. Right: illustration of one step in FlowCutter~\cite{flowcutter-journal, yang-wong-fbb}, where the smaller side (source-side) plus a piercing node are contracted to the corresponding terminal.
  }
\label{fig:flowconstructions}
\end{figure}

Sanders and Schulz~\cite{kaffpa} propose a flow-based refinement technique for bipartitions
that is scheduled on block pairs for $k$-way partitions.
The idea is to choose a subset $R$ of vertices around the cut of a bipartition $\Partition = \{V_1,V_2\}$.
A flow network is constructed by contracting $V_1 \setminus R$ to the source and $V_2 \setminus R$ to the sink.
The resulting maximum flow induces a possibly improved cut which may violate the balance~constraint.
Therefore, $|R|$ is chosen adaptively and a strongly-connected-component calculation is used to characterize the set of \emph{all} minimum cuts that can then be searched for a balanced bipartition.
Heuer \etal~\cite{kahypar-mf} generalize this approach to hypergraphs.
Gottesbüren \etal~\cite{reba-hfc, kahypar-hfc} accelerate flow-based hypergraph refinement by running the flow algorithm directly on the hypergraph.
By incorporating FlowCutter~\cite{flowcutter-journal, yang-wong-fbb}, which solves incremental flow problems to trade off cut size for better balance, they explore the solution space more effectively, which leads to partitions with smaller cuts/connectivity.
Examples are shown in Figure~\ref{fig:flowconstructions}.
FlowCutter~\cite{flowcutter-journal, yang-wong-fbb} is an iterative algorithm that starts with the minimum cut (and corresponding bipartition) between a given source and sink.
If the bipartition is imbalanced, the smaller block is grown by contracting all of its vertices plus an additional vertex (piercing node) to the corresponding terminal.
This vertex is chosen incident to the cut, and if possible without increasing the cut-size in the following iteration.
The resulting flow problems are nested in the sense that the terminal sets only grow, which makes the flow assignment from the previous iteration feasible but its maximality is violated.

\begin{figure}[t!]
\centering
\includegraphics[width=4.5cm]{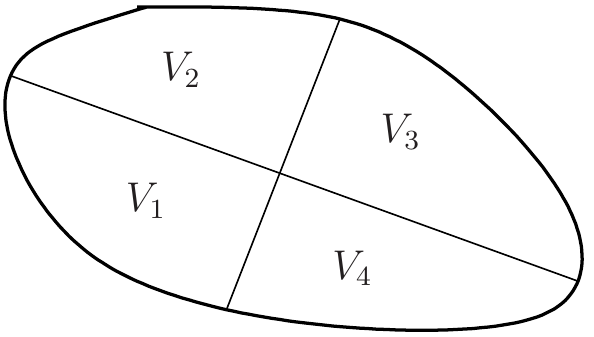}
\includegraphics[width=4.5cm]{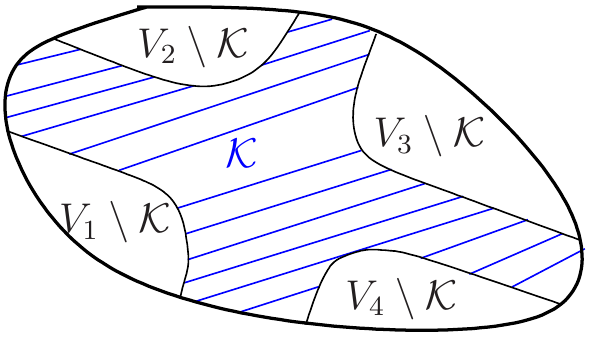}
\includegraphics[width=4.5cm]{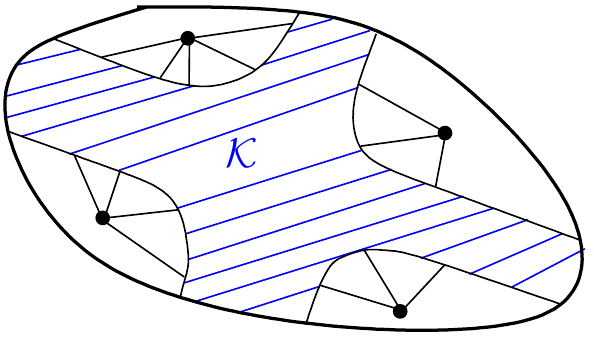}
\caption{ILP-based refinement. From left to right: a graph that is partitioned into four blocks; the set $\mathcal{K}$ close to the boundary that will stay in the ILP model; the model in which the sets $V_i \setminus \mathcal{K}$ have been~contracted. The latter is used as input to an ILP graph partitioning solver and the improved result is adopted. }
\label{fig:abstractexample}
\end{figure}

Henzinger~\etal~\cite{DBLP:journals/jea/HenzingerN020} use integer linear programming (ILP) to refine $k$-way partitions directly.
Analogously to flow-based refinement, a small region of vertices allowed to move is selected.
The remaining vertices are contracted to one super-vertex per block.
The ILP formulation is then run on this model graph.
An example is shown in Figure~\ref{fig:abstractexample}.
Techniques to speed up the ILP are giving a heuristic solution to the solver and symmetry breaking by
fixing sufficiently heavy super-vertices to their corresponding blocks.
Ugander and Backstrom~\cite{DBLP:conf/wsdm/UganderB13} use linear programming to select from a set of possible moves a
subset that yields the largest reduction in cut while satisfying the~balance~constraint.

\ifEnableExtendDONE
2013 Balanced Label Propagation for Partitioning Massive Graphs \cite{DBLP:conf/wsdm/UganderB13}  \\
2017 Engineering a Direct \emph{k}-way Hypergraph Partitioning Algorithm \cite{KaHyPar-K}\\
2019 Network flow-based refinement for multilevel hypergraph partitioning \cite{kahypar-mf} \\
2020 Advanced Flow-Based Multilevel Hypergraph Partitioning \cite{kahypar-hfc} \\
2020 (2018 conf version) ILP-Based Local Search for Graph Partitioning \cite{DBLP:journals/jea/HenzingerN020}

\fi

In experiments, flow based (hyper)graph refinement turns out to be a key ingredient of high quality partitioning, i.e., it has a much better cost-benefit-ratio than other approaches to improve quality like using ILPs, evolutionary techniques, restarts, V-cycles, etc.

\paragraph{Partitioning With Unevenly Distributed Vertex Weights}
Whereas real-world applications often use vertex and edge weights to accurately model the underlying problem,
the (hyper)graph partitioning research community commonly works with unweighted instances~\cite{HMS21, DBLP:phd/dnb/Schlag20}.
Multilevel algorithms incorporate techniques to prevent the formation of heavy vertices (restricting the
maximum allowed vertex weight or incorporate the weight of a vertex into the coarsening rating function as a
penalty term), but considerably struggle if such vertices are already present in the~input~\cite{HMS21},
as is the case for many real-world instances~\cite{ISPD98}.

There are two approaches to compute balanced $k$-way partitions under a tight balance constraint
in a multilevel partitioner: (i) ensure that initial partitioning finds a balanced partition~\cite{HMS21} and that refinement
applies node moves to the partition only when they satisfy the balance constraint, or (ii) allow intermediate balance
violations
\cite{DBLP:conf/iccad/DuttT97,DBLP:conf/aspdac/CaldwellKM00,CaldwellKM00,kaminpar}
and use rebalancing techniques to ensure that the final $k$-way partition is balanced
\cite{bipart,kaminpar,walshaw1997parallel,kabapEtr}.
Given that finding a balanced $k$-way partition is a NP-hard problem~\cite{garey1979computers} (reducible to the most
common version of the job scheduling problem), both approaches do not guarantee balance.

Recently, Heuer~\etal~\cite{HMS21} proposed a technique that enables partitioners
based on recursive bipartitioning to reliably compute balanced partitions in practice.
The idea is to preassign a small portion of heaviest vertices to one of the two blocks (treated as fixed vertices)
and optimize the objective function on the remaining vertices.

\subsection{Evolutionary Computation}%

A genetic or evolutionary algorithm (GA) starts with a population of individuals (in our case, partitions of the (hyper)graph) and evolves the population over several generational cycles or rounds.
If a genetic algorithm is combined with local search, then it is called a memetic algorithm (MA)~\cite{Kim11}.
In each round, the GA uses a selection rule to select good individuals based on the fitness of the individuals of the population and combines them to obtain improved offspring~\cite{goldbergGA89}. The combination step is typically done using a recombination operation.
When an offspring is generated, an eviction rule is used to select a member of the population to be replaced by the new offspring.
For an evolutionary algorithm, it is of major importance to preserve diversity in the population~\cite{baeckEvoAlgPHD96}; i.e., the individuals should not become too similar in order to avoid premature convergence of the algorithm.
This is usually achieved by using mutation operations and by using eviction rules that take similarity of individuals into account. All of the algorithms in the literature follow a rather generic overall scheme. Hence, we focus on the description of the recombination~operations.

Benlic and Hao \cite{DBLP:conf/ictai/BenlicH10}
cluster large sets of vertices together that have been assigned to same block in each individual to perform a recombination operation and combine their operator with tabu search.
The recombination operation is motivated by the observation that given a number of high-quality solutions, there is always a high number of vertices that are clustered together throughout these solutions.

Sanders and Schulz introduced a distributed evolutionary algorithm, KaFFPaE (KaFFPaEvolutionary) \cite{kaffpaE}.
The recombination operation uses a modified version of the multilevel graph partitioning solver within KaHIP \cite{kaffpa} that
will not contract edges that are cut in one of the input partitions. Thus, the better of the two input individuals can be used as initial partitioning and local search can efficiently exchange good parts of solutions on multiple levels of the multilevel hierarchy. Moreover, the recombination operation guarantees that the offspring is at least as good as the better of the two input individuals.

Ruiz and Segura~\cite{DBLP:journals/cys/RuizS18} present a memetic algorithm using a weighted matching-based recombination and diversity preservation that is similar to the algorithm by Benlic and Hao~\cite{DBLP:conf/ictai/BenlicH10}. More precisely, given two individuals, the recombination operation computes a matching in a bipartite graph where the left hand side represents the blocks of the first individual and the right hand side represent the blocks of the second individual. Edges between the vertices are weighted by the size of the intersection of the corresponding blocks. In this bipartite graph, a matching is computed and the corresponding intersections are used as the core of the sets in the new partition (offspring). Moreover, the algorithms use local search with negative cycle detection to further reduce the edge-cut of the computed partitions~\cite{kabapeE}.
When an offspring is inserted into the population, the individual having the highest similarity is evicted. Here, similarity between two individuals is based on the weight of the matching in the bipartite graph defined~above.

Henzinger~\etal~\cite{DBLP:journals/jea/HenzingerN020} provide a recombination operator that is based on integer linear programming and integrate this into KaFFPaE \cite{kaffpaE}.
The new recombination operation builds and solves an integer linear program from multiple individuals of the population.
Roughly speaking, the authors take $l$ individuals (partitions) and build an overlap graph by contracting pairs of nodes that are in the same block in every partition.
The original partitioning problem is then solved on the (much smaller) overlap graph using integer linear programming that is initialized with the block affiliations according to the partition that has the lowest cut value. When multiple partitions have the same cut value, one is chosen at random.

Andre~\etal~\cite{DBLP:conf/gecco/AndreS018} generalize KaFFPaE  to the multilevel memetic hypergraph
partitioner KaHyParE.
They also introduce a multilevel multi-point recombination operator that is applied to the best individuals.
This operator penalizes contraction of vertices that appear in nets that are cut in many parents.
Preen and Smith~\cite{DBLP:journals/tec/PreenS19} refine KaHyParE with an adaptive scheme to stop coarsening.%

Besides memetic algorithms for hypergraph and graph partitioning, there are also algorithms for other problem variations. For example, Moreira~\etal~\cite{DBLP:conf/gecco/MoreiraP018} and Popp~\etal\cite{DBLP:conf/alenex/PoppSSS21} gave memetic algorithms for acyclic (hyper)graph partitioning in which the input is an acyclic (hyper)graph and the partition has to fulfill an acyclicity constraint. Moreover, Schulz and Sanders~\cite{DBLP:conf/gecco/00010SW17} gave a distributed evolutionary framework to compute $k$-way vertex separators~in~graphs.

\ifEnableExtendDONE
\noindent 2011 An efficient memetic algorithm for the graph partitioning problem \cite{DBLP:journals/anor/GalinierBF11} \\
2011 An Effective Multilevel Memetic Algorithm for Balanced Graph Partitioning \cite{DBLP:conf/ictai/BenlicH10} \\
2012 Distributed Evolutionary Graph Partitioning\cite{kaffpaE} \\
2017 Distributed evolutionary $k$-way vertex separators \cite{DBLP:conf/gecco/00010SW17} \\
2018 Evolutionary multilevel acyclic graph partitioning \cite{DBLP:conf/gecco/MoreiraP018} \\
2018 Memetic multilevel hypergraph partitioning \cite{DBLP:conf/gecco/AndreS018} \\
2018 Memetic Algorithm with Hungarian Matching Based Crossover and Diversity Preservation \cite{DBLP:journals/cys/RuizS18} \\
2019 Evolutionary $n$-level Hypergraph Partitioning With Adaptive Coarsening \cite{DBLP:journals/tec/PreenS19}\\
2020 ILP-based local search for graph partitioning \cite{DBLP:journals/jea/HenzingerN020} (comb ILP + Evo)\\

\fi

\subsection{Streaming}
\ifEnableExtend

We consider a (hyper)graph partitioning algorithm as a streaming algorithm if it satisfies the following conditions: %
(i)~it receives as input the vertices or (hyper)edges of a graph in some sequential order,
(ii)~it cannot keep the whole (hyper)graph in memory (usually it can only spend linear memory on the number of vertices and blocks), and
(iii)~it performs permanent partitioning decisions on the fly based on partial information of the (hyper)graph.
More specifically, streaming partitioning algorithms usually follow the iterative load-compute-store sequence of operations shown in Figure~\ref{fig:streaming}.
This sequence of operations can be used to partition either a stream of vertices or a stream of (hyper)edges.
Moreover, it operates in batches that can contain either a single vertex/(hyper)edge or multiple of them at a time.

The most common streaming model is the \emph{one-pass} model, where
vertices are loaded one at a time alongside with their adjacency lists, then some logic is applied to permanently assign them to blocks.
The algorithm logic can be as simple as a Hashing function or as complex as scoring all blocks based on some objective and then assigning the vertex to the block with highest score.
There are other streaming models such as the sliding window model~\cite{patwary2019window}, where a fixed-size \emph{window} containing the next vertices is kept in memory to help with the one-pass decisions, and the buffered streaming model~\cite{jafari2021fast,DBLP:journals/corr/abs-2102-09384}, where \emph{buffers} or \emph{batches} of vertices are consecutively loaded, partitioned at once, and then permanently assigned to blocks.

Besides the traditional streaming graph partitioning approach in which vertices are assigned to blocks~\cite{nishimura2013restreaming,tsourakakis2014fennel,zhang2018akin,patwary2019window,awadelkarim2020prioritized,jafari2021fast,DBLP:journals/corr/abs-2102-09384}, many works on streaming graph partitioning adopt an edge-based approach ~\cite{bourse-2014,petroni2015hdrf,sajjad2016boosting,mayer2018adwise,hoang2019cusp}.
To disambiguate, we refer to this alternative approach as \emph{streaming edge partitioning} and we refer to the traditional approach either as streaming (hyper)graph partitioning or \emph{streaming vertex partitioning}.
In streaming edge partitioning, edges are loaded one by one and directly assigned to blocks on the fly in order to minimize the vertex-cut.
From an application perspective, streaming edge partitioning can be used to distribute the workload of an application represented by a graph where computations are edge-centric.
In this context, the vertex-cut of a partition indicates the number of vertices that will be \emph{replicated} or \emph{mirrored} in multiple blocks, which is usually associated with communication overhead.
Another common metric is the \emph{replication factor}, which is defined as the number of replicated vertices divided by the total of vertices in the graph.

\begin{figure}
	\centering
	\includegraphics[width=0.9\textwidth]{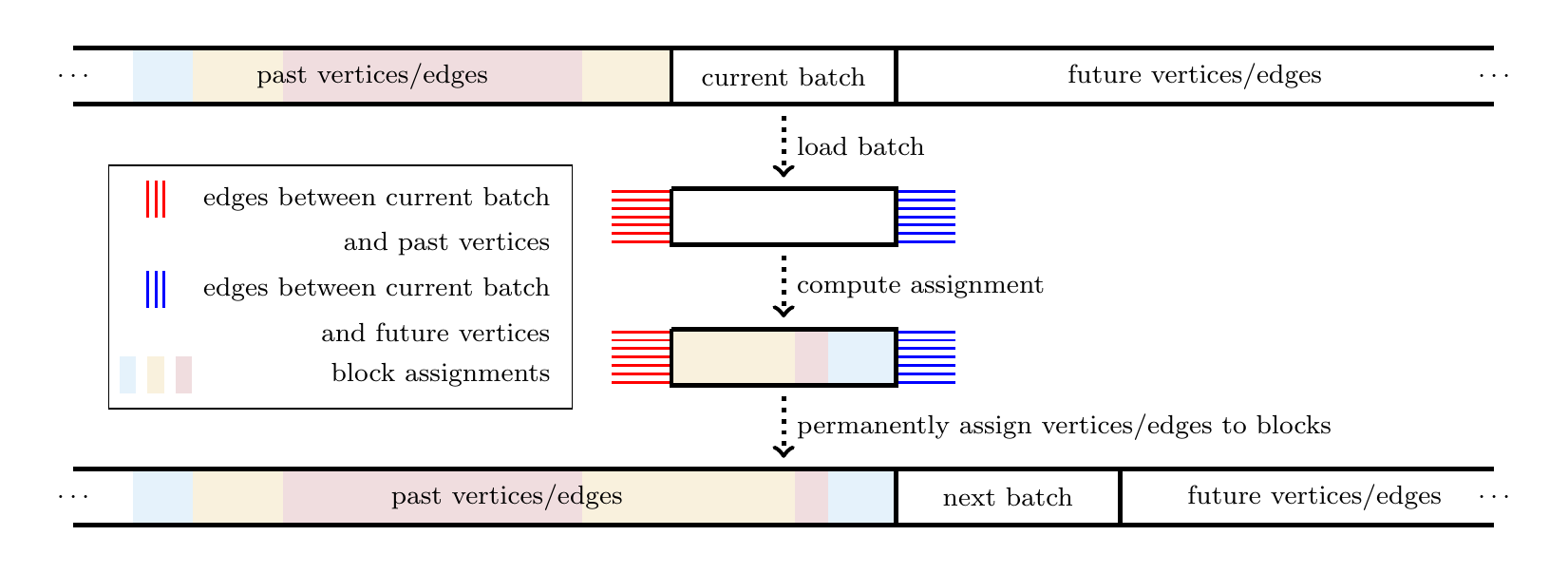}
	\vspace{-0.25cm}
	\caption{General iterative structure of streaming algorithms for graph partitioning.
}
	\label{fig:streaming}
	\vspace{-0.25cm}
\end{figure}

\subsubsection{One-Pass Streaming Vertex Partitioning}
Stanton and Kliot~\cite{stanton2012streaming}  propose heuristics to tackle the graph partitioning problem in the streaming model.
Among their most prominent heuristic is the one-pass method \emph{linear deterministic greedy}~(LDG) which produces solutions with the best overall edge-cut.
In this algorithm, vertex assignments prioritize blocks containing more neighbors and use a penalty multiplier to control imbalance.
Particularly, a vertex $v$ is assigned to the block $V_i$ that maximizes
$|V_i \cap \neighbors(v)|\Phi(i)$ with $\Phi(i)$ being a multiplicative degrading factor defined as $(1-\frac{|V_i|}{L_\text{max}})$
The intuition is that the degrading factor avoids to overload blocks that are already very heavy.
In case of ties on the objective function, LDG moves the vertex to the block with fewer vertices.
Overall, LDG partitions a graph in $O(m+nk)$ time.
Moreover, the authors also propose a simple one-pass methods based on \emph{hashing}, which has running time $O(n)$ and produces a poor~edge-cut.

Later, Stanton~\cite{stanton2014streaming} studies the streaming vertex partitioning problem from a more theoretical perspective.
The author proves that no algorithm can obtain an $o(n)$-approximation with a random or adversarial stream ordering.
Next, two variants of a randomized greedy algorithm are analyzed by using a novel coupling to finite Polya Urn~\cite{chung2003generalizations} processes, which intuitively explains the performance of the compared algorithms.

Tsourakakis~et~al.~\cite{tsourakakis2014fennel} propose Fennel,
a simple one-pass partitioning algorithm which adapts the widely-known clustering objective function \emph{modularity}~\cite{brandes2007modularity}.
Roughly speaking, Fennel assigns a vertex $v$ to a block $V_i$ in order to maximize an expression of type $|V_i\cap \neighbors(v)|-f(|V_i|)$, where $f(|V_i|)$ is an additive degrading factor.
More specifically, the authors defined the Fennel objective function by using
$f(|V_i|) = \alpha\gamma |V_i|^{\gamma-1}$, in which $\gamma$ is a free parameter and $\alpha = m \frac{k^{\gamma-1}}{n^{\gamma}}$.
After a parameter tuning made by the authors, Fennel uses $\gamma=\frac{3}{2}$, which implies $\alpha=\sqrt{k}\frac{m}{n^{3/2}}$.
Although the objective function penalizes imbalanced partitions, the authors define the possibility of a hard constraint to enforce balancing.
In their experiments, Fennel cuts fewer edges than LDG~\cite{stanton2012streaming} and, for some instances, it cuts roughly the same amount of edges as an offline partitioning algorithm.

Zhang~et~al.~\cite{zhang2018akin} propose AKIN, a streaming vertex partitioning algorithm for distributed graph storage systems.
AKIN is able to partition
graphs where the number of nodes $n$ is not known in advance by allowing the migration of vertices between blocks over time.
The assignment decisions are mainly based on the \emph{similarity} between vertices, which is evaluated with the \emph{Jaccard similarity coefficient}~\cite{hamers1989similarity}.
Given the (partial) neighborhoods of two vertices, this coefficient is defined as the ratio of their intersection over their union.
Initially, AKIN assumes a \emph{base block} for each vertex, which is given by a hash  function.
This base block is taken as a preliminary vertex assignment as well as a constant-time index for reaching information in the distributed graph storage.
More specifically, for each vertex, AKIN stores in its base block a fixed-length list containing its loaded neighbors with largest degree.
This list is used for computing the similarity between vertices.
As soon as an edge is loaded, AKIN assigns and migrates it and both of its endpoints to the block which maximizes a similarity-based heuristic. %
In the experimental evaluation, the version of AKIN keeping up to 100 neighbors per vertex cuts fewer edges compared to Fennel while maintaining equivalent imbalance and spending $10\%$ more running time.

\subsubsection{Restreaming Vertex Partitioning}
Nishimura~and~Ugander~\cite{nishimura2013restreaming} introduce a restreaming approach to partition vertices.
Their approach is motivated by scenarios where the same graph is
streamed multiple times.
In their model, a one-pass partitioning algorithm can pass multiple times through the entire input while the edge-cut is iteratively reduced.
The authors propose ReLDG and ReFennel, which are respective restreaming adaptations of linear deterministic greedy~(LDG)~\cite{stanton2012streaming} and Fennel~\cite{tsourakakis2014fennel}.
On the one hand, ReLDG modifies the objective of LDG to account only for vertex assignments performed during the current pass when computing block weights.
On the other hand, ReFennel uses the same objective as Fennel during restreaming, but its additive balancing degrading factor is increased after each pass in order to enforce balance.
Additionally, the authors prove that ReFennel converges after a finite number of restreams even without increasing the degrading factor.
Their experiments confirm that their restreaming methods can iteratively reduce edge-cut.

Awadelkarim and Ugander~\cite{awadelkarim2020prioritized} investigate how the order in which vertices are streamed influences one-pass vertex partitioning.
The authors introduce the notion of \emph{prioritized streaming}, where (re)streamed vertices are statically or dynamically reordered based on some predefined priority.
Their approach, which is a prioritized version of ReLDG, uses multiplicative weights of restreaming algorithms and adapts the ordering of the streaming process inspired by balanced label propagation.
In their experiments, the authors  consider a wide range of stream orderings.
The minimum overall edge-cut is obtained using a dynamic vertex ordering based on their own metric \emph{ambivalence}.
This approach is closely followed by a static ordering based on vertex degree.

\subsubsection{Buffered Streaming Vertex Partitioning}
Patwary et al.~\cite{patwary2019window} propose WStream, a simple streaming vertex partitioning algorithm that keeps a sliding window in memory.
The authors allow a few hundred vertices in the sliding window in order to obtain more information about a vertex before it is permanently assigned to a block based on a greedy function.
As soon as a vertex is allocated to a block, one more vertex is loaded from the input stream into the sliding window, which keeps the window size constant.
In their experiments, WStream cuts fewer edges than LDG and more edges than offline multilevel partitioning for most~tested~graphs.

Jafari et al.~\cite{jafari2021fast}
perform graph partitioning using a buffered streaming computational model.
The authors propose a shared-memory algorithm which repeatedly loads a batch of vertices from the stream input, partitions it using a multilevel scheme, and then permanently assigns the vertices to blocks.
Their multilevel %
scheme is based on a simplified structure where the one-pass algorithm LDG is used for coarsening, computing an initial partition, and refining it.
They parallelize LDG in a vertex-centric way by simply splitting vertices among processors, which yields a parallelization of the three steps of their multilevel scheme.
In their experiments, their algorithms cuts fewer edges than
LDG while scaling better than offline partitioning algorithms.

Faraj~and~Schulz~\cite{DBLP:journals/corr/abs-2102-09384} propose HeiStream, an algorithm which also partitions vertices in a buffered streaming model.
Their algorithm loads a batch of vertices, builds a graph model, and then partitions this model with a multilevel algorithm.
In their graph model, the vertices from previous batches assigned to each block are represented as a single big vertex fixed to the respective block.
Analogously,
an edge between a vertex $v$ from the current batch $b$ and a vertex $\bar{v}$ from a previous batch $\bar{b}$ is represented by an edge $(\bar{b},b)$.
In addition, when a vertex from a current batch has a neighbor from a future batch (\ie,  not yet streamed), their model compactly represents this neighbor in a contracted form.
Their multilevel algorithm has a traditional structure and components, except that the initial partitioning is a one-pass execution of Fennel and the label propagation refinement also optimizes the objective function used by Fennel to assign vertices to blocks.
In particular, the Fennel objective is extended to weighted graphs. %
In experiments, HeiStream cuts fewer edges than LDG, Fennel, and the buffered streaming algorithm proposed by Jafari~et~at.~\cite{jafari2021fast}, while being faster than Fennel for large numbers of blocks.

\subsubsection{Streaming Vertex Partitioning for Hypergraphs}
Alistarh~et~al.~\cite{alistarh2015streaming} propose Min-Max, a one-pass streaming
algorithm to assign the vertices of a hypergraph to blocks.
For each block, this algorithm keeps track of nets which contain pins in it.
This implies a memory consumption of $O(mk)$, which is more than the typical memory consumption of a streaming algorithm for graph partitioning.
When a vertex is loaded, Min-Max allocates it to the block containing the largest intersection with its nets while respecting a hard constraint for load balance.
The authors theoretically prove that their algorithm is able to recover a hidden \emph{co-clustering}
with high probability, where a co-clustering is defined as a simultaneous clustering of vertices and hyperedges.
In the experimental evaluation, Min-Max outperforms five intuitive streaming approaches with respect to load imbalance, while producing solutions up to five times more imbalanced than internal-memory algorithms such as~hMetis.

Ta{\c{s}}yaran~et~al.~\cite{tacsyaran2021streaming} propose improvements for the algorithm Min-Max~\cite{alistarh2015streaming}.
First, the authors present a modified version of Min-Max where, for each net, the blocks containing its pins are stored, instead of storing nets per block as done in Min-Max.
In their experiments, this modified version is three orders of magnitude faster than Min-Max while keeping the same cut-net but at the cost of up to two times more memory.
The authors also introduce three alternative adaptations using less memory than Min-Max: a constant upper-bound to restrict memory usage, the use of Bloom filters to answer membership queries, and the use of a hashing function to substitute the connectivity information between blocks and nets.
In their experiments, their three algorithms reduce the running time in comparison to Min-Max, especially the upper-bound and the hashing approaches, which are up to four orders of magnitude faster.
The memory consumption is also reduced in all three algorithms, especially the bloom filter and the hashing approaches, which respectively need one and two orders of magnitude less memory.
On the other hand, the three algorithms generate solutions with worse cut-net than Min-Max, especially the hashing approach, which increases the cut-net metric by up to an order of magnitude.
Moreover, the authors propose a buffered approach where a modified version of Min-Max goes multiple times through a buffer of vertices before permanently assigning them to blocks.
Their experimental results show that the buffered approaches improve partitioning quality in between $5\%$ and $20\%$ over Min-Max at the cost of a significant overhead in comparison with the fast modified versions of Min-Max.

\subsubsection{Streaming Edge Partitioning}

Bourse~et~al.~\cite{bourse-2014} study the streaming edge partitioning problem from a theoretical and practical perspective.
The authors start by providing two alternative formulations to vertex partitioning:
a \emph{basic} formulation whose objective is to minimize the traditional edge-cut metric and an \emph{aggregation} formulation whose objective is to minimize a modified edge-cut metric in which, for each vertex, only one incident cut edge is counted per block.
The authors provide two analogous formulations for edge partitioning assuming that each cut vertex is assigned to a \emph{master} block and replicated to the blocks containing its remaining edges.
In their \emph{basic} formulation, the objective is to minimize the number of edges incident to replicated vertices, \ie, edges not assigned to the master block of one of its endpoints.
In their \emph{aggregation} formulation, the objective is to minimize the number of replicated vertices.
Assuming a uniform random assignment of vertices and edges, the authors show that the expected
cost for aggregation-based edge partitioning is always smaller than or equal to the cost for aggregation-based vertex partitioning.
Moreover, it is proven that there is a one-to-one correspondence between vertex partitioning and edge partitioning in their basic formulation.
As a corollary, there exists an
approximate algorithm for edge partitioning which matches the best known approximation ratio for vertex partitioning.
Finally, the authors propose greedy heuristics for the streaming version of vertex partitioning and edge partitioning.
Their experiments demonstrate the practical advantages of edge partitioning over vertex partitioning and validate their approximation algorithms.

Petroni~et~al.~\cite{petroni2015hdrf} propose the ``High-Degree are Replicated First'' (HDRF) algorithm for streaming edge partitioning.
The HDRF algorithm, which aims at minimizing vertex-cut, operates by explicitly exploiting highly skewed power-law degree distributions for the edge assignment decision.
The intuition behind the algorithm is that high-degree vertices should preferably be cut, \ie, be replicated in several blocks,
since there are very few of them in power-law graphs.
The authors derive an average-case upper bound for its replication factor.
In their experiments, HDRF outperforms all the competitor algorithms with respect to replication factor.

Sajjad~et~al.~\cite{sajjad2016boosting} propose HoVerCut, a platform for streaming edge partitioning algorithms which can scale in multi-threaded and distributed systems by decoupling the state from the partitioning algorithm.
Each thread running HoVerCut receives a unique fraction of the streamed edges and partitions it using any streaming edge partitioning algorithm, such as HDRF~\cite{petroni2015hdrf}.
The threads keep \emph{local states} containing specific information required by the streaming algorithm, such as partial degree of processed vertices, partial block weights, and vertex assignments.
HoVerCut uses a buffering technique where these local states are shared on demand via a
\emph{global state} which is contained in a storage accessible by all threads.
All threads repeat the following sequence of steps asynchronously until the whole graph is partitioned: collect a buffer of edges, pull relevant information from global state, partition their buffers, and push local state to global state.
In their experiments with real-world and synthetic graphs, HoVerCut consistently speeds up the partitioning process of algorithms such as HDRF while roughly maintaining vertex-cut of the original partitioning algorithms.

Mayer~et~al.~\cite{mayer2018adwise} propose ADWISE, a window-based streaming algorithm for edge partitioning.
The algorithm keeps a dynamic window of edges and always picks the best edge in the window and assigns it to a block.
More specifically, the window starts containing only one edge; then its size dynamically increases or decreases in order to increase partition quality while roughly sticking to a given latency upper-bound.
Here, latency is defined as the time needed by the partitioning algorithm to assan edge in the current window to a block.
Thus ADWISE sticks to it by keeping the average latency to assign an edge under control throughout the execution.

Hoang~et~al.~\cite{hoang2019cusp} propose CuSP, a distributed and parallel streaming algorithm to partition edges based on user--defined policies.
The authors build CuSP as a programmable framework which is able to express the most common streaming edge partitioning strategies used in the literature.
The motivation for this approach is the wide variety of available partitioning policies and the lack of an integrated partitioner able to be deal with them.
In their experiments, CuSP can partition huge graphs with billions of vertices and hundreds of billions of edges in a few minutes, while allowing different partitioning policies.

\subsubsection{Experimental Studies}
Zainab~et~al.~\cite{DBLP:journals/pvldb/AbbasKCV18} provide a theoretical and experimental survey on the main one-pass streaming algorithms for vertex partitioning and edge partitioning.
After proposing a classification for these algorithms based on their requirements, formulations, and complexity, the authors experimentally evaluate them with respect to offline metrics such as imbalance, edge-cut, and vertex-cut.
Their results indicate that Fennel~\cite{tsourakakis2014fennel} achieves the lowest edge-cuts
while HDRF~\cite{bourse-2014} obtains the lowest replication factors.
For edge-cut partitioning algorithms, the authors also present experimental comparisons based on application metrics such as communication cost and total execution time.
In their experiments on application performance, HDRF~\cite{bourse-2014} and Greedy~\cite{gonzalez2012powergraph} yield higher partitioning cost that is compensated by lower computation time for iterative applications and better end-to-end latencies for single-pass stream processing applications.
Hashing, on the other hand, has the lowest partitioning time, which increases the overall performance of applications in which data locality does not impact communication cost considerably.

Pacaci~and~{\"O}zsu~\cite{pacaci2019experimental} present an experimental survey of streaming algorithms for vertex and edge partitioning.
They investigate performance, resource usage and scalability of the algorithms.
The authors study both vertex and edge-partitioning algorithms based on offline graph metrics and online graph queries.
Pacaci and {\"O}zsu conclude
that primarily minimizing edge-cut or vertex-cut fails to represent the workload performance in practice,
as some workload characteristics and graph structures are not addressed by these metrics.
For example, all tested algorithms suffer from poor workload distribution in practice even when the algorithms theoretically should provide balanced partitions.
Furthermore, their experiments show that edge-cut and vertex-cut are reliable indicators of network communication, although this does not always improve the application performance in practice.
On the other hand, experiments also show that hashing is an effective strategy for scale-out graph databases since it provides a good trade-off between throughput and latency for online graph queries.

\fi

\subsection{Repartitioning}
\ifEnableExtend

Graph repartitioning is an extension of graph partitioning which copes with dynamic graphs, \ie, graphs whose set of vertices and edges are modified over
time.
Assuming that a graph is partitioned, dynamic changes on its components can make its blocks imbalanced and its edge-cut or vertex-cut larger.
Figure~\ref{fig:repartitioning} illustrates the evolution of a graph over time followed by repartitioning efforts,
In general, a rough upper bound for repartitioning quality can be obtained by partitioning the whole graph from scratch every time it is modified.
Since this is expensive, many repartitioners use faster~approaches.

\begin{figure}
	\centering
	\includegraphics[width=0.9\textwidth]{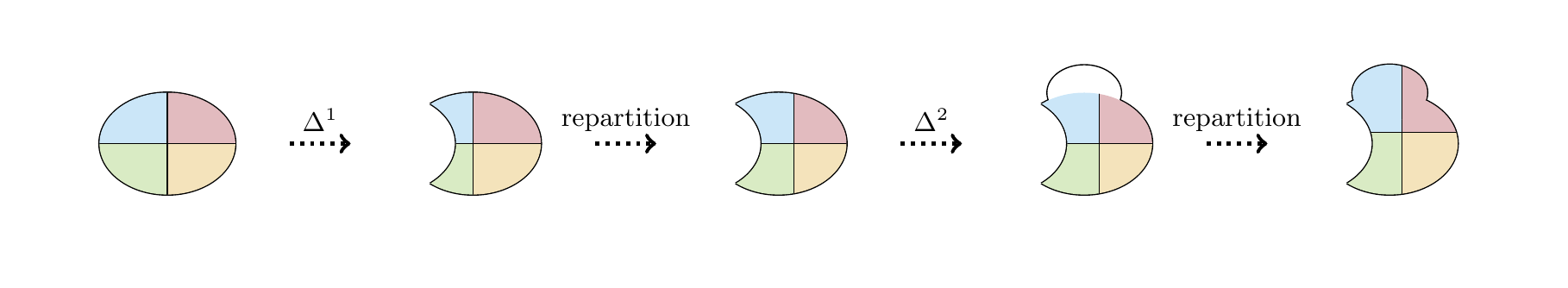}
	\vspace{-0.25cm}
	\caption{Illustration of the repartitioning process. A  graph which has an initial $4$-way partition goes through two subsequent changes: $\Delta^{1}$ where portions of the graph are removed, $\Delta^{2}$ where new elements (vertices and edges) are inserted in the graph. After each of these changes, a repartitioning routine adapts the previous $4$-way partition to the new graph state while optimizing for one or more metrics such as edge-cut, imbalance, and migration of elements between blocks.
	}
	\label{fig:repartitioning}
	\vspace{-0.25cm}
\end{figure}

Vaquero~et~al.~\cite{vaquero2014adaptive} propose a distributed repartitioning algorithm for large-scale graph processing systems.
In their algorithm, hashing is used as initial partitioning for all vertices.
Based on local information, a label propagation scheme iteratively migrates vertices to blocks where most of their neighbors are located until a convergence is achieved.
This procedure natively deals with modifications in the graph.
Their experiments show that their approach is able to keep a better edge-cut and application performance upon graph changes in comparison to the hashing algorithm.
The authors did not compare their algorithm against other state-of-the art repartitioning~algorithms.

Xu~et~al.~\cite{xu2014loggp} propose LogGP, a repartitioning algorithm with centralized coordination for graph processing systems.
Their algorithm analyzes and reuses historical information to improve the partitioning.
Particularly, LogGP builds a hypergraph by combining the graph with hyper-edges which represent previous historical partitions.
This hypergraph is then partitioned by a novel streaming pin partitioning algorithm.
During the execution, the system uses statistical inferences from running logs to optimize the partitioning process.
In their experiments, LogGP outperforms state-of-the-art streaming and repartitioning algorithms with respect to edge-cut and application~runtime.

Nicoara~et~al.~\cite{nicoara2015hermes}
propose a lightweight repartitioning algorithm and integrate it in their own distributed social network database system Hermes.
Within Hermes, the initial partitioning is performed with hashing or Metis, while their repartitioning algorithm is triggered when there is a change in the vertex set of the graph.
Repartitioning uses iterative local improvement of balance and edge cut. The associated data is moved only when the improvement process is finished.
In their experiments with real-world workloads, their algorithm produces roughly the same edge-cut while migrating up to an order of magnitude less data in comparison with Metis (repartitioning from scratch).
Moreover, their algorithm improves around $1.7$ times over hashing with respect to aggregate throughput.

Huang~and~Abadi~\cite{huang2016leopard} propose Leopard, an algorithm that solves graph repartitioning while also replicating vertices.
Leopard can potentially take advantage of any one-pass vertex partitioning algorithm.
This is possible since these algorithms are vertex-centered, so they can be used locally on a dynamic graph at any moment.
Moreover, Leopard is the first algorithm to integrate vertex replication with the repartitioning.
This replication provides fault tolerance and improves locality, which even increases the partitioning quality, since it is mostly based on local information.
Their experiments show that Leopard without vertex replication maintains a partition quality comparable to running Metis from scratch at any moment.
On the other hand, Leopard with vertex replication further reduces edge-cut by up to an order of magnitude.

Kiefer~et~al.~\cite{kiefer2016penalized} propose a repartitioning algorithm optimized for systems whose performance does not scale linearly with the input size.
The authors start by formulating a penalized version of the graph partitioning problem by defining the weight of a block as the sum of weights of its vertices plus a penalty function which grows monotonically with the cardinality of the block.
In practice, this formulation penalizes resource consumption, which models the nonlinear behavior of the performance of some real-world systems.
The authors modify a static multilevel algorithm for graph partitioning to solve the penalized version of the problem, then this multilevel algorithm is used within an adapted version of a hybrid repartitioning strategy.
Their hybrid repartitioning algorithm works by triggering a local refinement whenever the balance constraint is violated and computing a whole new partition in the background after a given number of refinements.
More specifically, this new partition is computed from scratch; afterward its blocks are mapped onto the blocks of the previous partition to minimize migration costs (see Section~\ref{subsubsec:process_mapping} for more details on process mapping). Then it replaces the current partition only if the new edge-cut compensates the migration overhead.
The authors implement their algorithms on the Metis~\cite{karypis1998fast} framework and show that their static algorithm takes $28\%$ more time than Metis on average while keeping a linear complexity on the amounts of vertices, edges and blocks.
Finally, their repartitioning algorithm is able to keep imbalance, edge-cut, and migration time low throughout~graph~updates.

Fan~et~al.~\cite{fan2020incrementalization} study the problem of repartitioning vertices (edges) in order to simultaneously minimize edge-cut (vertex-cut) and the modifications to the initial partition.
The authors prove robust intractability even for restricted cases of
the problem.
Moreover, they provide nontrivial proofs that this problem is unbounded with respect to edge-cut and vertex-cut.
Based on the presented theorems, the authors derive a strategy to convert successful partitioning algorithms into repartitioning algorithms while ensuring small migration cost and keeping the same partition quality bounds of the original algorithm.
As proof of concept, the authors apply their strategy on algorithms such as Fennel~\cite{tsourakakis2014fennel} (vertex partitioning) and HDRF~\cite{bourse-2014} (edge partitioning).
Finally, the authors validate their algorithm~experimentally against the repartitioning algorithm Hermes~\cite{nicoara2015hermes} and  algorithms to partition from scratch such as ParMetis~\cite{parmetis-conference}.
In their experiments, their algorithms are up to an order of magnitude faster then their original counterparts while retaining a comparable or even better partition quality.

\fi

\subsection{Objective Functions and Problem Variations}\label{ss:variations}
The simple balanced (hyper)graph partitioning problems studied in most of this survey does not capture all aspects relevant to applications. Therefore, additional issues such as multiple constraints, directed edges, matrix partitioning, process mapping, and various ways of modelling communication costs have been considered.
\subsubsection{Objective Functions.}

Kaya~\etal~\cite{DBLP:conf/ppam/KayaUC13} present an experimental study on the
influence
of different partitioning models and metrics on the performance of
parallel sparse matrix-vector multiplication (SpMxV).
To do so, real-life and artificial sparse matrices are partitioned
using several matrix partitioning models. The authors then measure the running time
of matrix-vector multiplications using three different SpMxV
implementations and determine the influence of different partition
metrics on the running time by performing a regression analysis.
The experimental evaluation shows that  \emph{the right} partition model and metric
depends on the number of processing elements used, the specific SpMxV implementation
and the size and structure of the matrix.
More specifically,  minimizing
the total number of messages sent is more important than minimizing
the per-processing element send volume, in particular if the number of processing elements is $\ge 64$.
The checkerboard partition
model~\cite{DBLP:journals/siamsc/CatalyurekAU10} reduces most of the
partition metrics and generally achieves the lowest running time.

Originally proposed by Schloegel~\etal~\cite{DBLP:conf/europar/SchloegelKK99},
the multi-objective approach attempts to minimize multiple objective
functions simultaneously.
To this end, Deveci~\etal~\cite{DBLP:journals/jpdc/DeveciKUC15} recently
introduced the multi-objective hypergraph partitioner UMPa.
While previous work~\cite{DBLP:journals/siamsc/UcarA04, Bisseling2005} on this problem variation minimized multiple objectives
in multiple phases, where the first phase optimizes the primary objective
which may not be worsened by later phases, UMPa uses a single-phased approach
to optimize a primary metric in combination with a secondary and
tertiary metric.
This is done by greedily moving boundary vertices such that the primary
metric is optimized, while using the secondary (and tertiary) metrics
for tie-breaking if multiple eligible target blocks maximize the
primary (or secondary) metric.
The choice of the best target block of a boundary vertex depends on its
gain values, i.e., the change of the metric if the vertex is moved to
another block.
To this end, the authors describe how gain values can be computed
efficiently for the communication volume and number of messages
metrics as well as both metrics on a per-block basis.

\subsubsection{Multi-constraint Graph Partitioning}
Balanced graph partitions are usually constrained by an upper limit on
the weight of each block.
In the multi-constraint graph partitioning model introduced by
Karypis~and~Kumar~\cite{DBLP:conf/sc/KarypisK98}, each vertex of the
graph is associated with a vector of weights, and a balance constraint
is imposed on each weight.
Multi-constraint partitioning was extended for hypergraphs~\cite{Catalyurek01-SC},
enabling 2D coarse-grain partitioning of sparse matrices/graphs~\cite{DBLP:journals/siamsc/CatalyurekAU10}.
Recently, Recalde~\etal~\cite{DBLP:journals/ejco/RecaldeTV20}
introduced an exact algorithm for the multi-constraint graph partitioning problem.
Two integer programming formulations are provided, and the authors prove
several families of inequalities associated with the respective
polyhedras.  %
Using a branch-and-bound based approach, Recalde~\etal can solve real-world
instances with \numprint{30} vertices in approximately half a minute of CPU
time.
Moreover, several of the proven families of
inequalities significantly reduce the number of branch-and-bound vertices
and the optimality gap.

\subsubsection{Directed Acyclic (Hyper)graph  Partitioning}
If the input instance is a directed acyclic graph (DAG), one is often
interested in finding a partition with an acyclic quotient graph.
Moreira~\etal~\cite{DBLP:journals/heuristics/MoreiraPS20} show
that perfectly balanced graph partitioning is NP-complete even
under the additional acyclicity constraint, and that there are no
constant-factor approximation algorithms for $k \ge 3$.
Herrmann~\etal~\cite{DBLP:journals/siamsc/HerrmannOUKC19, DBLP:conf/ccgrid/HerrmannKUKC17}
and Moreira~\etal~\cite{DBLP:journals/heuristics/MoreiraPS20} both
propose multilevel heuristics for DAG partitioning.
Popp~\etal~\cite{DBLP:conf/alenex/PoppSSS21} adapt these techniques
to directed acyclic hypergraph partitioning.

Moreira~\etal~\cite{DBLP:journals/heuristics/MoreiraPS20} coarsen the
graph using size-constraint label propagation~\cite{DBLP:journals/heuristics/MeyerhenkeS016}.
Since this algorithm is not adapted to DAGs, contracting the computed
clusters creates coarse graphs with cycles.
Thus, the partitioner computes an initial acyclic partition
\emph{before} coarsening.
Restricting clusters to single blocks then ensures that the partition
of the input graph can also be used as a partition of the coarsest
graph.
Herrmann~\etal~\cite{DBLP:journals/siamsc/HerrmannOUKC19} present an
acyclic clustering algorithm which ensures that coarse graphs are
acyclic.
The top-level of a DAG vertex $v$ is defined as the
longest distance from any source of the DAG to $v$.
Based on the top-level of adjacent vertices, conditions
for inter- and intra-cluster edges are formulated and it is shown that a clustering
respecting those conditions yields an acyclic coarse graph.
Moreira~\etal compute an initial partition by cutting the
topologically ordered vertices of the graph into $k$ chunks of equal
size.
Herrmann~\etal adapt greedy graph growing to initial
bipartitioning~\cite{DBLP:conf/ccgrid/HerrmannKUKC17} and initial
$k$-way partitioning~\cite{DBLP:journals/siamsc/HerrmannOUKC19}.
Moreover,  bipartitioning the graph as if it was
undirected (using, for instance, Metis~\cite{karypis1998fast}) and fixing the
acyclicity constraint afterwards~\cite{DBLP:journals/siamsc/HerrmannOUKC19} seems promising.
For refinement, both Herrmann~\etal and Moreira~\etal adapt the FM
algorithm to DAG partitioning by restricting the set of possible vertex
moves to those that do not violate the acyclicity constraint.
Herrmann~\etal~\cite{DBLP:journals/siamsc/HerrmannOUKC19} note that
this condition can be checked efficiently if the partition has only
two blocks.

\subsubsection{Symmetric Rectilinear Matrix Partitioning}
Ya\c{s}ar~\etal~\cite{DBLP:journals/corr/abs-1909-12209, DBLP:journals/corr/abs-2009-07735}
consider the symmetric rectilinear sparse matrix partitioning problem.
Here, the columns and rows of a sparse matrix are partitioned using
the same partition vector (in contrast to the more general rectilinear matrix
partitioning problem \cite{DBLP:conf/para/ManneS96, DBLP:conf/irregular/GrigniM96}, in which different vectors are used).
This yields a tiling of the sparse matrix with square tiles on the
diagonal.
The goal is to find a partition that minimizes the weight
(i.e., the sum of all nonzero matrix entries assigned to a tile)
of the heaviest tile or, given an
upper limit on the weight of a tile, to find a partition subject to
the constraint with a minimum number of cuts.
The authors show that both problem variations are
NP-complete~\cite{DBLP:journals/corr/abs-2009-07735}, propose
heuristics to optimize either objective, %
and provide efficient implementations thereof.
Running time is reduced by sparsifying the
matrix.
Using these techniques, the algorithm can partition a graph representing the Twitter social network
with approximately 1.5 billion edges in less than 3 seconds on a
modern~24 core system. %

\subsubsection{Process Mapping.}
\label{subsubsec:process_mapping}

\ifEnableExtend
Distributing a set of parallel and communicating processes among the processing elements~(PEs) of a high-performance system is one of the main applications of graph partitioning.
Assume these processes and their communication are respectively represented by the vertices and edges of an \emph{application graph}, which can be partitioned in order to split processes among PEs.
When different pairs of PEs communicate with each other at different speeds, the edge-cut does not reflect the total communication cost.
Process mapping is a generalization of graph partitioning which aims at partitioning a set of processes into blocks of roughly equal size and mapping the blocks onto PEs in order to minimize the \emph{total communication cost} or related objective functions.

The exact total communication cost of a given mapping of processes onto PEs depends on a combination of \emph{bandwidth} and \emph{latency}, which depends on a multitude of factors in practice.
Computationally, the total communication cost is usually modeled based on an (implicit) \emph{topology matrix} which contains the distances between PEs.
Given a mapping of an application graph onto PEs, the total communication cost is defined as the sum of weights of the cut edges multiplied by the distance between the PEs containing their respective endpoints.
The \emph{hierarchical topology} is a special case where PEs are contained in a multi-layered hierarchy of modules and sub-modules.
Latency-based topologies are another special kind of topology in which PEs are contained in a \emph{topology graph} such that the distance between a pair of PEs is defined as the number of edges in the shortest path between them.
The total communication cost in latency-based topologies is also known as the \emph{Coco(.)} or \emph{hop-byte} objective function.
An alternative objective function to the total communication cost is the bandwidth-based metric \emph{maximum congestion}, which is defined as the maximum number of message exchanges through any link of the topology graph.
Another possible objective is the \emph{maximum dilation}, which is defined as the maximum communication cost directly associated with a pair of PEs for a given mapping.

The algorithms for process mapping can be categorized in two groups.
On the one hand, the single-phase algorithms
combine process mapping with graph partitioning  \cite{DBLP:journals/fgcs/WalshawC01,DBLP:conf/hpcn/PellegriniR96, hierarchprocessmap}, such that the objective of the partitioning -- commonly edge-cut -- is typically replaced by a function that measures communication cost.
On the other hand, the two-phase algorithms decouple partitioning and mapping \cite{schulz2017better,brandfass2013rank,heider1972computationally,muller2013optimale,glantz2015algorithms}.
In a two-phase algorithm, a default graph partitioning algorithm is used to partition a communication graph into $k$ blocks while typically minimizing edge-cut.
Afterwards, the quotient graph of the partitioned communication graph is mapped onto PEs in order to minimize the total communication cost or other objective function.
The reader is referred to~\cite{GPOverviewBook,SPPGPOverviewPaper} for older works on~process~mapping.

Deveci~et~al.~\cite{DBLP:conf/ipps/DeveciKUC15} propose a greedy
construction algorithm and two refinement algorithms to map processes to PEs based on a topology graph.
The construction algorithm iteratively picks the process with highest connectivity to the already mapped processes.
Then, this process is assigned to the module which minimizes hop-bytes, which is found by performing a breadth-first search on the topology graph.
Both their refinement methods are based on process swapping, but one of them aims at minimizing hop-bytes while the other one aims at minimizing
congestion.
The authors experimentally compare their algorithms on a torus-based system against the default mapper of the system as well as Scotch~\cite{Scotch} and LibTopoMap~\cite{Walshaw07}.
Their algorithms induce performance increases of $43\%$ and $23\%$ for a communication-only application and a sparse matrix vector multiplication, respectively, compared to the default mapper of the system, and
have the best overall results among the competitors.

Deveci~\etal~\cite{8666156} also propose a scheme that exploits geometric
partitioning to map processors to PEs.  The application data and
PEs are partitioned separately using the MultiJagged geometric partitioner~\cite{DBLP:journals/tpds/DeveciRDC16}.
Blocks from the data partition are then mapped to
corresponding blocks in the PE partition, effectively assigning interdependent
data to ``nearby'' PEs in the network.  The method is appropriate for
applications that have geometric coordinates as well as graph data.

Vogelstein~et~al.~\cite{vogelstein2015fast} solve an application-agnostic problem where there are two graphs with the same number of vertices which are bijectively
mapped onto one another in order to minimize the number of induced edge disagreements.
The authors call this problem \emph{graph matching problem} and formulate it as a quadratic assignment problem~(QAP).
They solve it with a non-linear approximation algorithm based on the gradient vector.
Their algorithm has complexity $O(k^3)$ and performs better than the previous state-of-the-art~\cite{umeyama1988eigendecomposition,singh2007pairwise,zaslavskiy2008path} regarding running time and objective function for over $80\%$ of the QAPLIB benchmark library~\cite{burkard1997qaplib}.

Glantz~et~al.~\cite{glantz2015algorithms}
propose two greedy algorithms to bijectively map blocks onto PEs assuming a communication graph.
Their most successful algorithm is an adaptation of the greedy mapping algorithm proposed
by M{\"u}ller-Merbach~\cite{muller2013optimale}.
The algorithm by M{\"u}ller-Merbach~\cite{muller2013optimale} initially computes the total communication volume for each process and also the sum of distances from each PE to all the others.
Afterwards, it iteratively assigns the process with largest communication volume to the PE with the smallest total distance.
Glantz~et~al.~\cite{glantz2015algorithms} modify this algorithm by scaling the distance with the amount of communication to be done.
This modification improves the overall mapping quality with respect to the quality measures maximum congestion and maximum dilation.

Glantz~et~al.~\cite{glantz2018topology} propose a local improvement algorithm for one-to-one process mapping in which the hardware topology is a partial cube, \ie, an isometric subgraph of a hypercube.
The authors exploit the regularity of these topologies to label PEs as well as processes with bit-strings along convex cuts.
These bit-strings permit fast computation of distances between PEs and the implementation of effective
hierarchical refinement methods to improve the mapping induced by the labels.
Their experimental results show that their algorithm reduces the total communication cost of the mappings produced by state-of-the-art mapping algorithms ~\cite{Scotch,glantz2015algorithms} in a range from $6\%$ to $34\%$ while the mapping time stays within the same order of magnitude.

Schulz~and~Träff~\cite{schulz2017better} solve the process mapping for a hierarchical topology using a two-phase approach.
Their algorithm uses the state-of-the-art partitioner KaHIP~\cite{kahipWebsite} for the partitioning phase and then a multi-section for the one-to-one mapping phase.
The referred multi-section algorithm recursively partitions the quotient graph of the previously obtained partition throughout the layers of the communication topology in a top-down direction.
After this algorithm obtains a one-to-one mapping, a light-weight modification of
the swap-based refinement method proposed by Brandfass~et~al.~\cite{brandfass2013rank} is executed to further minimize the communication cost.
Their experiments show that their approach produces mappings with lower total communication cost in comparison with alternative approaches which combine the same KaHIP-based first phase with state-of-the-art algorithms for one-to-one mapping~\cite{muller2013optimale,Walshaw07,glantz2015algorithms}.
Kirchbach~et~al.~\cite{kirchbach2020better} further improve the two-pass approach by Schulz~and~Träff~\cite{schulz2017better}.
In particular, the authors specialize the partitioning phase to apply KaHIP multiple times, namely throughout the hierarchical topology.
Kirchbach~et~al experimentally show that their best algorithm is faster while also decreasing the total communication cost in comparison to the algorithm from Schulz~and~Träff~\cite{schulz2017better}.

Faraj~et~al.~\cite{hierarchprocessmap} propose a single-phase multilevel algorithm to solve the process mapping for a hierarchical topology.
Their multi-section scheme uses a top-down recursive multi-section algorithm to the initial mapping.
This initial mapping is then improved during the uncoarsening using modified versions of label propagation, $k$-way FM, multi-try FM, and quotient graph refinement which are specifically designed to minimize the total communication cost.
Additionally, their algorithm implicitly represents the hierarchical topology in a condensed binary labeling of PEs, which uses linear space in the number of PEs and provides the distance between PEs in constant time.
Experimentally, their algorithm improves the total communication cost over the algorithms proposed by Schulz~and~Träff~\cite{schulz2017better} and Kirchbach~et~al.~\cite{kirchbach2020better} while also having a lower mapping~time.

Predari~et~al.~\cite{DBLP:journals/corr/abs-2107-02539} propose a single-phase distributed algorithm to solve the process mapping for a hierarchical topology.
The authors model the system hierarchy as an implicit labeled tree and label the vertices of the communication graph in order to implicitly induce the mapping.
The algorithm optimizes the mapping by using an
adapted version of parallel label propagation where the labeling scheme is used for quick gain computations.
In their experiments, the proposed algorithm has good scalability for up to thousands of PEs while achieving a total communication cost smaller than the distributed algorithms ParMetis~\cite{parmetis-conference} and ParHIP~\cite{DBLP:journals/tpds/MeyerhenkeSS17} and comparable cost to the state-of-the-art sequential mapping algorithm proposed by Faraj~et~al.~\cite{hierarchprocessmap}.

Kirchbach~et~al.~\cite{von2020efficient}
address the process mapping problem assuming processes that communicate in a sparse stencil pattern and PEs organized in Cartesian grids.
First, the authors prove that the Cartesian mapping is already NP-hard for a
two-dimensional Cartesian grid and a one-dimensional stencil.
Then three fast construction algorithms which exploit the regularity of the problem are proposed.
Their algorithm with best overall results is the \emph{stencil strips} algorithm, which partitions the grid into strips of lengths close to the scaled length of an optimal bounding rectangle of the coordinates of the target stencil.
In their experimental evaluation using the MPI\_Neighbor\_alltoall routine of MPI,
their algorithms are up to two orders of magnitude faster than general purpose graph mapping tools such as VieM~\cite{viemaWebsite} while resulting in a similar communication performance.
Moreover, their algorithms are up to three times faster than other Cartesian grid mapping algorithms while resulting in a much better communication performance.

\fi

\section{Parallel Algorithms}
\label{sec:parallelalgos}

Most of the initial work on graph partitioning involved sequential algorithms.
These algorithms have been extended to work in distributed memory environments,
in particular for balancing processor workloads in parallel applications.
In this use case, a distributed memory application already has a
distribution of the graph; for memory scalability, the entire graph is
not stored in every processor.
Thus, distributed-memory partitioning algorithms do not typically have a
global view of the entire graph;
they often make partitioning decisions based on partial views of
local graph data. As a result, they can have lower solution quality than their
sequential counterparts.  Still, distributed memory algorithms are crucial
for graphs that are too large to fit in a single memory space, and for
applications wishing to partitioning their data dynamically to adjust for
changing computational workloads.

In recent years, progress has been made on shared memory algorithms as
shared memory architectures offer greater flexibility than distributed memory
architectures.
For example, random memory accesses or atomic updates can be done orders of magnitude faster compared to distributed memory machines.
Because shared-memory algorithms have a global view of the graph,
they can achieve the same solution quality as their sequential predecessors.
They are not feasible, however, for extremely large graphs
that do not fit in a single memory space.

\subsection{Shared Memory}

Most of the algorithms discussed in this section are multilevel algorithms.
Therefore we structure the discussion into the separate contributions in each phase.
The multilevel algorithms we consider are Mt-Metis by Lasalle and Karypis~\cite{mt-metis}, Mt-KaHiP by Akhremtsev \etal~\cite{mt-kahip-journal}, Mt-KaHyPar by Gottesbüren \etal~\cite{mt-kahypar} (hypergraphs), KaMinPar by Gottesbüren \etal~\cite{kaminpar} (deep multilevel), and BiPart by Maleki \etal~\cite{bipart}.
The following work is focused on transferring the sequential approaches to shared memory as faithfully as possible, while sometimes incorporating lessons learned from distributed memory.

\subsubsection{Coarsening}
Most coarsening algorithms are based on greedy matching or greedy clustering.
These algorithms visit vertices in some order (e.g., random), calculate the ratings (e.g., heavy-edge) for joining neighboring clusters (or match with neighbor), and then join the highest rated cluster.
Visiting vertices in parallel yields faithful parallelizations of the sequential approaches.
The challenge is to prevent inconsistent clustering decisions between threads.

{\c{C}}ataly{\"{u}}rek \etal~\cite{ParallelPaToH} propose parallel schemes for agglomerative clustering and greedy matching that use locking, in addition to a lock-free version of greedy matching that resolves conflicts in a second pass.
The lock-based algorithms first try to lock the visited vertex, calculate ratings, and then iterate through the candidates.
When a better candidate is found, its lock is tested.
If successful, the old candidate is replaced and unlocked.
For the lock-free resolution scheme, the matches are stored in a global array $M$ without protecting write access.
After one pass, vertices $u$ with $M[M[u]] \neq u$ are matched with themselves $M[u] \gets u$.
These are vertices whose match was visited concurrently and matched to a different vertex.
In their evaluation, the lock-based algorithms fare equally well as the sequential versions in terms of cardinality and heavy-edge metric, whereas the resolution-based one falls off by 10\% in the heavy-edge metric.

The resolution-based matching scheme is also used by LaSalle and Karypis~\cite{mt-metis} in Mt-Metis.
On inputs with skewed degree distributions (such as complex networks), maximal matchings are small and thus coarsening converges too slowly.
If too few vertices are matched after a pass, pairs of non-adjacent vertices that have identical neighbors and low degree are matched.
If still too few vertices are matched, any vertex pair that shares a common neighbor can be matched.
This technique is dubbed two-hop matching~\cite{opt-mt-metis}.

Akhremtsev \etal~\cite{mt-kahip-journal} instead use parallel size-constrained label propagation clustering~\cite{DBLP:conf/wea/MeyerhenkeSS14} to coarsen skewed inputs more effectively.
No locks are employed, so cyclic cluster joins may occur.
The cluster size constraint ensures that initial partitioning can find a feasible partition.
Cluster sizes are updated with atomic instructions.
The size constraint is checked before the update, but this offers no atomic consistency.
Hence, the instructions' results are checked to guarantee the size constraint (and revert if exceeded).
For coarsening it is not necessary to strictly adhere to the size constraint (whereas it is for refinement to guarantee balance).
Label propagation is also used in KaMinPar~\cite{kaminpar}, but the atomic check is omitted for coarsening (not for refinement).

For Mt-KaHyPar, Gottesbüren \etal~\cite{mt-kahypar} use parallel agglomerative clustering but defer the locking after the rating and target cluster selection, locking only the visited vertex and target cluster.
Vertices trying to cyclically join each other are detected and resolved on-the-fly, recursively merging the associated clusters.

\subsubsection{Initial Partitioning}

For initial partitions, LaSalle and Karypis~\cite{mt-metis} use recursive bipartitioning.
Each thread sequentially computes a bipartition, from which the best one is selected.
For recursion, the threads are statically split into two groups, working on the two separate subgraphs.
This is later improved~\cite{opt-mt-metis} to threads cooperating on one bipartition if the graph is sufficiently large instead of independent sequential trials.
Akhremtsev \etal~\cite{mt-kahip-journal} compute independent $k$-way partitions with sequential KaHiP~\cite{kaffpa}.
The downsides of these two approaches are potentially severe load imbalance (recursive bipartitioning) or being a sequential bottleneck.
To overcome this, Gottesbüren \etal~\cite{mt-kahypar} use a work-stealing task scheduler instead of splitting threads for recursive bipartitioning with parallel (un)coarsening.
For flat bipartitions, a portfolio of sequential algorithms is run independently in parallel~\cite{KAHYPAR-IP, DBLP:conf/alenex/SchlagHHMS016}.

The deep multilevel approach~\cite{kaminpar} offers additional parallelism through repetitions (forks) at different coarsening levels.
At these stages the graphs may be too small to make efficient use of parallel (un)coarsening, such that splitting threads off for repetitions comes at little to no running time penalty.
Using randomized components (coarsening, initial bipartitioning, refinement) thus offers multiple diversified solutions from which the best is chosen.

Slota \etal~\cite{pulp} use parallel $k$-source BFS to compute flat $k$-way partitions.
Vertices join the blocks of their parents.
Their approach does not use coarsening, which justifies using a flat parallel algorithm at this stage.

Maleki \etal~\cite{bipart} parallelize greedy hypergraph growing~\cite{karypis1998fast, PaToH} for bipartitioning.
All vertices are assigned to block $V_0$.
Then the gains of moving the vertices in block $V_0$ to $V_1$ are computed.
The $\sqrt{|V|}$ highest rated vertices are moved, before the gains of vertices remaining in block $V_0$ are recomputed.
This is repeated until the bipartition is balanced.

\subsubsection{Refinement}\label{sss:parRefinement}

The two most crucial questions that must be addressed in the refinement phase are how to maintain a balanced partition and how to make sure that concurrent moves do not result in worse cuts.
Some refinement algorithms are more difficult to parallelize than others.
Unsurprisingly, the more sophisticated techniques, in particular those able to escape local minima, are more difficult to parallelize, as intermediate negative gain moves are required.
In particular FM and KL are known to be P-complete~\cite{PHARD}, which makes the existence of poly-log depth algorithms~unlikely.

\paragraph{Label Propagation}

The most straightforward approach to parallelize is label propagation: simply visit vertices in parallel.
While gains can be incorrect due to concurrent moves in their neighborhood, this does not affect cut optimization too much in practice.
Balance can be ensured by using atomic instructions to update block weights~\cite{mt-kahip-journal, mt-kahypar}, or by first collecting moves and approving in a second step~\cite{socialhash, bipart, mt-kahypar-det-tr}.
During refinement,
it is important to enforce the size constraint by checking the result of the atomic instruction, in order to guarantee a balanced partition.

\paragraph{Greedy}

LaSalle and Karypis~\cite{mt-metis} parallelize Metis' greedy refinement by statically assigning vertices to threads and running the sequential algorithm on the local vertices.
The sequential algorithm performs FM on boundary vertices,
but stops once no positive gain moves remain.
This eases parallelization since no moves must be reverted, and thus there is no serial move order to observe.
However, it does eliminate the ability to escape local minima.
Moves are communicated to other threads of neighboring vertices via message buffers.
The threads frequently check their buffers to update local gains.
Further, the refinement is split into an upstream and a downstream pass, where vertices are only allowed to move into blocks with higher (resp. smaller) identifier than their current block.
This avoids accidental cut-degrading concurrent moves~\cite{parmetis-conference} but restricts the search space since some blocks quickly become close to overloaded.
Balance checks are done optimistically with locally updated block weights.
Since this may result in balance violations, synchronization points after passes are used to revert some moves to restore balance.

\paragraph{Parallel Localized FM}

Localized FM~\cite{kaffpa} is a variant of FM that starts with a few boundary vertices and expands its search space to neighbors of moved vertices.
It is good at escaping local minima due to allowing negative gain moves, but not wasting too much time on unpromising areas through short search sprints.
This approach can be parallelized by performing multiple independent localized searches, as opposed to standard FM which is difficult to parallelize efficiently~\cite{PHARD}.
Akhremtsev \etal~\cite{mt-kahip-journal} organize the boundary vertices in a queue that is randomly shuffled.
Threads repeatedly poll seed vertices from the queue and perform localized FM around their seeds.
Moves are not communicated to other threads.
Instead, each thread maintains a local partition -- an array for block weights and a hash table for partition IDs.
The rationale for keeping moves private is that reverting (negative gain) moves at the end of a localized search confuses other searches.
Searches may overlap in their local vertices, but each vertex is moved only once (atomic test-and-set).
Once the global queue is empty, the local move sequences of the threads are concatenated into a single sequence.
To ensure a balanced partition and no cut degradation, the gains of this sequence are recomputed sequentially, and the prefix that yields the smallest edge cut subject to the balance constraint is applied.

In Mt-KaHyPar, Gottesbüren \etal~\cite{mt-kahypar} apply local move sequences to the global partition as soon as a local optimum is found.
This provides more accurate information to the other threads.
Additionally, the resulting move order more adequately reflects the partition state on which the move decisions of the localized searches are based.
Finally, it reduces the memory for hash tables, while keeping the negative gain moves at the end of a localized search private.
This is important for hypergraphs where there is more information
(e.g., $\pinsinpart(e,i)$)
stored in local partition data structures than in graph partitioning.
Moves applied to the global partition are rejected if they violate the balance constraint and block weights are maintained with atomic instructions.

\paragraph{Parallel Flow-based Refinement}
The flow-based refinement from Section~\ref{lab:flowbasedrefinment} offers two parallelism sources that are investigated in~\cite{mt-kahypar-flows-tr}.
The core routine works on bipartitions, such that it can be applied to different block pairs of a $k$-way partition in parallel.
A basic version only runs non-overlapping block pairs in parallel, but the authors show that overlapping searches are feasible with certain restrictions, and necessary for better parallelism.
The second parallelism source is the flow algorithm, where the well-known push-relabel algorithm is nicely amenable to parallelization.
The approach is integrated in the state-of-the-art parallel multilevel framework Mt-KaHyPar.
Experiments show that the partition quality of the new algorithm is on par with the highest quality sequential code (KaHyPar), while being an order of magnitude faster when using 10 threads.

\paragraph{Hill Scanning}

LaSalle and Karypis extend their greedy refinement to escape local minima to some extent~\cite{mt-metis-hill-climbing} by employing a simplified variant of localized search.
Whenever only negative gain moves remain in the thread-local priority queue, a small group of vertices around $u$ (up to 16) is incrementally constructed, with the hope that moving the entire group reduces the cut.
The group is constructed in the same way as localized FM expands its search; however, the expansion stops as soon as an improvement is possible.
The selected vertices are not restricted to the thread-local ones, as opposed to the greedy algorithm.
One restriction is that all vertices must be moved to the same block.

\paragraph{Techniques for Accurate Gains}

Gottesbüren \etal~\cite{mt-kahypar} double-check gains for localized FM and label propagation using a technique named \emph{attributed gains}.
Recall that $\pinsinpart(e, i)$ denotes the number of pins that hyperedge $e$ has in block $V_i$.
When vertex $u$ is moved from block $V_s$ to $V_t$, $\pinsinpart(e, s)$ is (atomically) decremented and $\pinsinpart(e, t)$ is incremented for each $e \in I(u)$.
Reducing $\pinsinpart(e, s)$ to zero attributes an $\omega(e)$ connectivity reduction, whereas increasing $\pinsinpart(e, t)$ to one attributes an $\omega(e)$ increase to moving $u$.
If the overall attributed gain of a move is negative, it is reverted.

The global moves in localized FM are collected in a sequence in the order in which they were applied.
As in Mt-KaHiP, Gottesbüren \etal perform a gain recalculation on this sequence in Mt-KaHyPar, for which a parallelization is proposed~\cite{mt-kahypar}.
Analogously to attributed gains the last pin of a hyperedge $e$ moved out of a block is attributed an $\omega(e)$ reduction (if no pin moved in before) and the first pin moved into the block is attributed an $\omega(e)$ increase (if none were contained before).
This can be calculated in two passes over the pins of $e$.
Each hyperedge incident to a moved vertex is handled independently in parallel, with gain attributions distributed using atomic fetch-and-add.

Finally, the authors~\cite{mt-kahypar} show that updating a global gain table~\cite{fiduccia1982lth, FromPQs, KaHyPar-K} can be performed in parallel with atomic instructions.
Analogously to attributed gains, gain table updates are triggered by specific observed values when the $\pinsinpart(e, i)$ values are updated.
For neighbors of moved vertices their priority queue key is updated with the associated table entries.
Further, when a vertex is extracted from the priority queue, its key is checked against the table entries, and reinserted with the new key, if worse.
This way, searches gradually update their priority queues to global partition changes without resorting to message passing.

\paragraph{Rebalancing}

If vertices are moved concurrently, it does not suffice to check whether each single move would preserve balance, but rather some synchronization mechanism is necessary.
Therefore, some refinement algorithms cannot guarantee balanced partitions, so there is a need for explicit rebalancing algorithms.
Furthermore, even if some refinement algorithms can guarantee balance, intermediate balance violations can lead to smaller cuts after a rebalancing step.

Slota \etal~\cite{pulp} use label propagation with gains multiplied by $L_{\max} / |c(V_i)|$ to favor moving into lighter blocks~\cite{DBLP:conf/icde/MartellaLLS17}.
Maleki \etal~\cite{bipart} use the same approach as their graph growing parallelization for rebalancing two-way partitions, starting with a non-empty lighter block that is gradually filled.

Lasalle and Karypis~\cite{mt-metis} integrate rebalancing into their parallel greedy refinement.
The thread-local move buffers are traversed in reverse order and moves into overloaded blocks are reverted.
Each thread is responsible for restoring excess weight proportional to the amount it moved into that block.

For large $k$, Gottesbüren \etal~\cite{kaminpar} use one priority queue per overloaded block with the ratio of highest gain (to a non-overloaded block) and vertex weight as key.
Each queue is filled with just enough vertices to remove the excess weight, but neighbors of moved vertices in the same former block are inserted, for the case that designated target blocks become close to overloaded.
Parallelism is achieved by emptying different overloaded blocks in parallel.

Gottesbüren \etal~\cite{mt-kahypar} allow controlled balance violations in the gain recalculation step of localized FM in Mt-KaHyPar.
Label propagation on the next level is often able to rebalance, and explicit rebalancing similar to label propagation is employed on the finest level.

\paragraph{Parallel $n$-level (Un)Coarsening}

Gottesbüren \etal~\cite{MT-KAHYPAR-Q} propose an $n$-level version~\cite{kaspar, KaHyPar-K} of Mt-KaHyPar~\cite{mt-kahypar}.
With sequential $n$-level, only one vertex is (un)contracted at a time, which is inherently sequential but offers high solution quality through highly localized refinement.
In the parallel version, vertex pairs are contracted asynchronously with vertices to be contracted organized in a hierarchical forest data structure.
The forest yields precedence conditions (bottom up) for the asynchronous contractions.
With this, vertices in different subtrees can be contracted independently in parallel, as soon as their children are finished.
Fine-grained locking is employed to edit the hypergraph data structure, and dynamically maintain consistency of the forest.

Uncoarsening introduces parallelism by uncontracting batches of $b > 1$ vertices in parallel.
The batches are constructed by traversing the forest in top-down order, assembling contractions that can be reverted independently in a batch.
Uncontracting a batch resolves the last dependencies required to uncontract the next
batch. The vertices of the current batch serve as seeds for highly localized parallel refinement (label propagation and FM).

\paragraph{Determinism}

Researchers have advocated the benefits of deterministic parallel algorithms for
several decades~\cite{DBLP:conf/popl/Steele90, DBLP:conf/ppopp/BlellochFGS12}, including ease of debugging, reasoning about performance,
and reproduciblity.
A downside is less flexibility in terms of algorithm design choices and potential overheads for synchronizing optimization decisions.
For example, the deterministic version of Mt-KaHyPar is a factor of 1.16 slower than an equivalent non-deterministic configuration, and exhibits a factor of 1.029 higher connectivity (both aggregated using geometric mean).
Interestingly this degradation stems from coarsening~\cite{mt-kahypar-det-tr}, not refinement.

Maleki \etal~\cite{bipart} propose BiPart~\cite{bipart}, a deterministic multilevel recursive bipartitioning algorithm.
For coarsening, vertices are matched to their smallest hyperedge with ID hashes as tie breakers.
All vertices matched to the same hyperedge are contracted without restricting coarse vertex weights.
Initial partitioning uses the parallel greedy graph growing described above.
For refinement, the gain for moving each vertex to the opposite block is computed in parallel.
Let $l_0, l_1$ denote the number of vertices with positive gain in block $V_0, V_1$.
Then the $\min(l_0, l_1)$ vertices with highest gains are swapped, an approach that was already proposed for Social Hash~\cite{socialhash}.
If the hypergraph has unit vertex weights this maintains a balanced partition.
However this algorithm is used within the multilevel framework, where non-uniform vertex weights occur from coarsening.
Thus, the graph growing rebalancing described above is employed.
Note that gains become stale after a neighbor is moved, and thus the overall cut reduction~is~not the~sum~of~the~gains.

Gottesbüren and Hamann~\cite{mt-kahypar-det-tr} present a deterministic version of Mt-KaHyPar~\cite{mt-kahypar}.
The label propagation style algorithms for preprocessing, coarsening and refinement of Mt-KaHyPar are re-implemented in a deterministic fashion using synchronous local moving~\cite{SLM}.
Move decisions do not depend on other moves in the current round.
Rounds are further split into sub-rounds to incorporate more up-to-date information.
After each sub-round some of the moves are approved and some are denied, for example due to the balance constraint or a cluster size constraint for coarsening.
For refinement, the same method as BiPart~\cite{bipart} and SocialHash~\cite{socialhash} is used: swapping highest gain prefixes between block pairs.
To incorporate non-unit vertex weights, the cumulative weights of all prefixes of the move sequences are computed.
The best prefix combination is then selected similar to a parallel merge.

\subsection{Distributed-Memory}\label{subsec:dist}
Distributed (hyper)graph partitioning is one way to handle large inputs that do not fit into the main memory
of a single machine. In the distributed memory model, several processors (PEs) are interconnected via a
communication network, and each has its private memory inaccessible to others. Computational tasks on each PE usually
operate independently only on local data representing a small subset of the input. Intermediate computational results must be
exchanged via dedicated network communication primitives.

Distributed (hyper)graph processing algorithms require that the vertices and edges of the input are partitioned among the
processors. Since many applications use balanced (hyper)graph partitioning to obtain a good initial assignment, much simpler
techniques are used in distributed partitioners.
There exist range-based~\cite{DBLP:journals/tpds/MeyerhenkeSS17} and
hash-based partitioning techniques~\cite{xtrapulp,xtrapulp-journal}. The former splits the vertex IDs into equidistant ranges,
which are then assigned to the PEs.
Since both techniques do not consider the (hyper)graph structure,
it could lead to load imbalances or high communication overheads.
However, one could also migrate vertices as more information about the structure of the (hyper)graph is available,
e.g., when recursing on a subgraph obtained via recursive bipartitioning~\cite{berger1987partitioning, DBLP:journals/tpds/DeveciRDC16}.
If geometric information are available, one can also use space-filling curves~\cite{DBLP:conf/icpp/LoozTM18,DBLP:journals/algorithms/AydinBM19}.

Each PE then stores the vertices assigned to it and the edges incident to them.
The edges stored on a PE can be incident to local vertices or vertices on other PEs (also called \emph{ghost} or \emph{halo} vertices).
We say that a PE is adjacent to another PE if they share a common edge.
Processors must be able to identify adjacent PEs to propagate updates, e.g. if we move a vertex to a different block,
we have to communicate that change to other PEs in the network such that local search algorithms can work
on accurate partition information. However, each communication operation introduces overheads that can limit the
scalability of the system. Thus, the main challenge in distributed (hyper)graph partitioning is keeping the
global partition information on each PE in some sense up to date while simultaneously minimizing the
required communication.

The remainder of this section describes the algorithmic core ideas of recent publications in that field and abstracts
from the physical placement of the vertices and the actual representation of the distributed (hyper)graph data structure.
However, we assume that each vertex knows on which PE its neighbors are stored.

\paragraph{Local Search.}
The label propagation heuristic is the most widely-used local search algorithm in
distributed systems~\cite{DBLP:journals/tpds/MeyerhenkeSS17,socialhash,DBLP:conf/icde/MartellaLLS17,xtrapulp,
karypis1996parallel,walshaw1997parallel,DBLP:conf/ipps/TrifunovicK04,Zoltan}. Other approaches
schedule sequential $2$-way FM~\cite{fiduccia1982lth} on adjacent block pairs in parallel~\cite{kappa,chevalier2008pt,KirmaniR13}.
However, this limits the available parallelism to at most the number of blocks $k$.

Parallel label propagation implementations mostly follow the \emph{bulk synchronous parallel} model.
In a computation phase, each PE computes for its local vertices their desired target block.
In the communication phase, updates are made visible to other PEs via personalized all-to-all communication~\cite{DBLP:journals/tpds/MeyerhenkeSS17,xtrapulp}.
Meyerhenke~\etal~\cite{DBLP:journals/tpds/MeyerhenkeSS17} use an asynchronous communication model.
If the computation phase of a PE ends, it sends and receives updates to and from other PEs
and immediately continues with the next round.

In the parallel setting, the move gain of two adjacent vertices may suggest
an improvement when moved individually, but moving both simultaneously may
worsen the solution quality. Therefore, some partitioners use a vertex coloring~\cite{karypis1996parallel}
or a two-phase protocol where in the first phase, vertices can only move from a block $V_i$ to $V_j$ if $i < j$
and vice versa in the second phase~\cite{DBLP:conf/ipps/TrifunovicK04,Zoltan,mt-metis}.
Many systems do not use any techniques to protect against move conflicts.
This can be seen as an optimistic strategy assuming that conflicts rarely happen in practice.

The \emph{Social Hash Partitioner}~\cite{socialhash} (Facebook's internal hypergraph partitioner) also uses
the label propagation heuristic to optimize $\text{fanout}(\Pi) := \frac{1}{|E|} \sum_{e \in E} \lambda(e)$
where $\Pi = \{V_1,\ldots,V_k\}$ is a $k$-way partition.
The authors note that the label propagation algorithm can easily get stuck in local optima
for fanout optimization and suggest a probabilistic version of the fanout metric, called
$\text{p-fanout}(\Pi) := \frac{1}{|E|} \sum_{e \in E} \sum_{V_i \in \Pi} 1 - (1  - p)^{\pinsinpart(e, V_i)}$ for some probability $p \in (0,1)$.
The probabilistic fanout function samples the pins of hyperedges with probability $p$ and represents the expected fanout for a
family of similar hypergraphs. Thus, it should be more robust and reduce the impact of local minima.

Other recently published distributed local search techniques are based on vertex swapping techniques that preserve the balance of the partition.
Rahimian~\etal~\cite{DBLP:conf/saso/RahimianPGJH13} present JA-BE-JA that uses such an approach.
The algorithm iterates over the local vertices of each PE and for each vertex, it considers all adjacent
vertices as swap candidates. If no partner was found, it selects a random vertex from a sample as a candidate.
If the selected vertex is assigned to a different PE, the instantiating PE sends a request with all the required information such that
the receiving PE can verify whether or not the swap operation would improve the edge cut. On success, both vertices change their blocks.
Additionally, simulated annealing is used to avoid local minima.

Aydin~\etal~\cite{DBLP:journals/algorithms/AydinBM19} implement a distributed partitioner that computes
a linear ordering of the vertices, which is then split into $k$ equally-sized ranges to obtain an initial $k$-way partition.
The idea is similar to space-filling curves~\cite{DBLP:books/daglib/0035382, DBLP:journals/tpds/PilkingtonB96},
but does not require geometric information.
The initial ordering is computed by assigning labels to a tree constructed via agglomerative
hierarchical clustering. Afterward, it sorts the labels of the leaves to obtain an initial ordering.
To further improve the ordering, it solves the \emph{minimum linear arrangement} problem that tries to optimize $\sum_{(u,v) \in E} |\pi(u) - \pi(v)|\omega(u,v)$
where $\pi(u)$ denotes the position of $u \in V$ in the current ordering. To do so, it uses a two-stage MapReduce algorithm that is repeated until
convergence: First, each vertex computes its desired new position as the weighted median of its neighbor's positions. Second, the final positions
are assigned to the vertices by resolving duplicates~with~simple~ID-based~ordering.
The second local search algorithm performs vertex swaps.
It splits the vertices of each block into $r$ disjoint sets and pairs adjacent blocks of the partition and randomly their sets.
The paired sets are then mapped to the processors that perform the following algorithm:
It sorts the vertices in both sets according to their cut reduction if moved to the opposite block and
swaps the vertices with the highest combined cut reduction.

\paragraph{Balance Constraint.}
The label propagation algorithm only knows the exact block weights at the beginning of each computation phase.
In the computation phase, block weights are only maintained locally. In the communication phase, the combination
of all moves may result in a partition that violates the balance constraint. Thus, partitioners based on this scheme have
to employ techniques to ensure balance.

The distributed multilevel graph partitioner ParHIP~\cite{DBLP:journals/tpds/MeyerhenkeSS17} divides a label propagation
round into subrounds and restores the exact block weights with an All-Reduce operation after each subround. Note that this
does not guarantee balance but gives a good approximation of the block weights when the number of moved vertices in a subround
is small.

Slota~\etal~\cite{xtrapulp} implemented a distributed graph partitioner that alternates between
a balance and refinement phase, both utilizing the label propagation algorithm. In
the refinement phase, each PE maintains approximate block weights $a(V_i) := c(V_i) + \gamma \Delta(V_i)$
where $c(V_i)$ is the weight of block $V_i$ at the beginning of the computation phase, $\Delta(V_i)$ is the weight of vertices
that locally moved out respectively into block $V_i$, and $\gamma$ is a tuning parameter that depends on the number of PEs. Each PE
then ensures locally that $a(V_i) \le L_{\max}$ for all $i \in \{1, \ldots, k\}$.
In the balancing phase, the gain of moving a vertex to block
$V_i$ is multiplied with $\frac{L_{\max}}{a(V_i)}$. As a consequence, moves to underloaded blocks become
more attractive. In a subsequent publication~\cite{xtrapulp-journal}, the approach is generalized to the multi-constraint
partitioning problem, where each vertex is associated with multiple weights.

Recently, probabilistic methods were proposed that preserve the balance in expectation~\cite{socialhash,DBLP:conf/icde/MartellaLLS17}.
The Social Hash Partitioner~\cite{socialhash} aggregates the number of vertices $S_{i,j}$ that want to move from block $V_i$ to $V_j$
after each computation phase at a dedicated master process.
Then, a vertex part of block $V_i$ is moved to its desired target block $V_j$ with probability
$\frac{\min(S_{i,j}, S_{j,i})}{S_{i,j}}$. This ensures that the expected number of vertices that move from block $V_i$ to $V_j$ and vice versa is the same and,
thus, preserves the balance of the partition in expectation. However, each PE moves its highest ranked vertices with probability one and all remaining
probabilistically.
Martella~\etal~\cite{DBLP:conf/icde/MartellaLLS17} moves a vertex $u$ to its desired target block $V_j$ with
probability $\frac{L_{\max} - c(V_j)}{M_j}$ where $M_j$ are the number of vertices that want to move to block $V_j$.
The advantage of the probabilistic method is that only the number of vertices preferring a different block need to be communicated
instead of all moves.

\paragraph{Multilevel Algorithms.}
Although it is widely known that multilevel algorithms produce better partitions than flat partitioning schemes,
the systems used in industry, e.g., at Google~\cite{DBLP:journals/algorithms/AydinBM19} or
Facebook~\cite{socialhash,DBLP:conf/icde/MartellaLLS17}, are primarily non-multilevel algorithms.
The main reason for this is that the scalability of multilevel algorithms is often limited to a few hundred processors~\cite{kappa,KirmaniR13}.
Furthermore, most parallel multilevel systems implement matching-based coarsening
algorithms~\cite{DBLP:journals/jpdc/WalshawCE97,chevalier2008pt,KirmaniR13,kappa,DBLP:conf/ipps/TrifunovicK04,Zoltan}
that are not capable to efficiently reducing the size of today's complex networks (power-law node degree distribution).
The most prominent distributed multilevel algorithms are
Jostle~\cite{DBLP:journals/jpdc/WalshawCE97}, ParMetis~\cite{parmetis-conference}, PT-Scotch~\cite{chevalier2008pt},
KaPPa~\cite{kappa}, ParHIP~\cite{DBLP:journals/tpds/MeyerhenkeSS17} and ScalaPart~\cite{KirmaniR13} for graph, and
Parkway~\cite{DBLP:conf/ipps/TrifunovicK04} and Zoltan~\cite{Zoltan} for hypergraph partitioning.

Meyerhenke~\etal~\cite{DBLP:journals/tpds/MeyerhenkeSS17} build the parallel multilevel partitioner ParHIP
that uses a parallel version of the size-constraint label propagation algorithm~\cite{DBLP:conf/wea/MeyerhenkeSS14}.
The algorithm is used to compute a clustering in the coarsening phase
and as a local search algorithm in the refinement phase. To obtain an initial partition of the coarsest graph, it uses
the distributed evolutionary graph partitioner KaFFPaE~\cite{kaffpaE}.
On complex networks, ParHIP computes edge cuts $38\%$ smaller than those of
ParMetis~\cite{parmetis-conference} on average, while it is also more than a factor of two faster.

Wang~\etal~\cite{DBLP:conf/icde/WangXSW14} use a similar approach that also utilizes the label propagation algorithm
to compute a clustering in the coarsening phase. The algorithm is implemented on top of Microsoft's Trinity graph engine~\cite{shao2012trinity}.
The partitioner additionally uses external memory techniques to partition large graphs on a small number of machines.
However, it does not perform multilevel refinement (the initial partition is projected to the input graph).

\begin{figure}
\includegraphics[width=8cm]{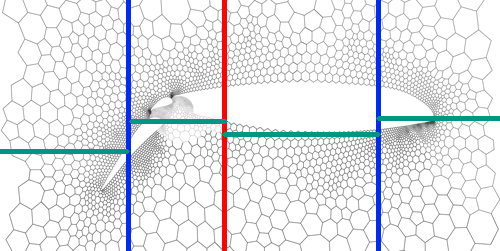}
\caption{A visualization of an airfoil. The graph has coordinates for each vertex. Geometric partitioning algorithms use this type of information for partitioning. In the example above, recursive coordinated bisection (always split the graph along the shorter axis) has been applied to derive a partition.}
\label{fig:airfoildual}
\end{figure}

\paragraph{Geometric Partitioners}
Many graphs are derived from geometric applications and are enriched with
coordinate information (e.g., each vertex is associated with a $d$-dimensional point). A mesh with coordinate information and a partition based on this coordinates is shown in Figure~\ref{fig:airfoildual}.
Geometric partitioning techniques use this information to partition the corresponding point set
into $k$ equally-sized clusters while minimizing an objective function defined on the clusters.
The objective function should be chosen such that it implicitly optimizes the desired graph
partitioning metric (e.g., sum of the lengths of all bounding boxes approximates the total communication
volume~\cite{DBLP:journals/tpds/DeveciRDC16}).
Since geometric methods ignore the underlying structure of the (hyper)graph, the quality of the partitions
is often inferior compared to traditional multilevel algorithms. However, these algorithms are often simpler leading
to faster and more scalable algorithms.
Prominent techniques use space-filling curves~\cite{DBLP:books/daglib/0035382, DBLP:journals/tpds/PilkingtonB96}
that maps a set of $d$-dimensional points to an one-dimensional line.
A fundamental property of this curve is that points that are close in the original space are also
close on the line. Other approaches recursively divide the space via cutting planes such as
Octree-based partitioning~\cite{minyard1998octree},
recursive coordinate bisection~\cite{berger1987partitioning,simon1991partitioning}, and
recursive inertial bisection~\cite{taylor1994study,DBLP:journals/concurrency/Williams91}.  The
MultiJagged algorithm of Deveci~\etal~\cite{DBLP:journals/tpds/DeveciRDC16} uses
multisectioning rather than bisection to reduce the depth of the recursion
and speed computation relative to recursive coordinate bisection; its hybrid
implementation uses MPI and Kokkos~\cite{kokkos} to support both distributed memory message passing between PEs
and multithreading or GPU computation within PEs.

Recently, Von Looz~\etal~\cite{DBLP:conf/icpp/LoozTM18} presented a scalable balanced $k$-means algorithm to partition geometric graphs.
The $k$-means problem asks for a partition of a point set $P$ into $k$ roughly equally-sized clusters such that the squared distances of each point
to the mean of its cluster is minimized (in the following also referred to as the center of a cluster).
Clusters obtained with this problem definition tend to have better shapes than computed with previous methods
and also produce better partitions when measured with graph metrics~\cite{DBLP:journals/pc/MeyerhenkeMS09}.
They present a parallel implementation
of Lloyd's greedy algorithm~\cite{lloyd1982least} that repeats the following steps until convergence. First, each point $p \in P$ is assigned to the
cluster that minimizes the distance of $p$ to its center. Afterwards, the center of each cluster is updated by calculating the arithmetic mean of all points
assigned to it. To achieve balanced cluster sizes, an influence factor $\gamma_c$ is introduced individually for each cluster $c$ and
$\text{eff\_dist}(p, \text{center}(c)) := \text{dist}(p, \text{center}(c)) / \gamma_c$ is used as the distance of point $p$ to the center of a cluster $c$.
If a cluster $c$ becomes overloaded, the influence factor $\gamma_c$ is decreased and otherwise it is increased.
Thus, underloaded clusters become more attractive. The implementation replicates the cluster centers and
influence factors globally and after each computation phase, it updates the values via a parallel sum operation. To obtain an initial solution, it sorts the
points according to the index on a space-filling curve and splits the order into $k$ equally-sized clusters. Furthermore, it establishes a lower bound for the distances
of each point to the second-closest cluster which allows to skip expensive distance computations for most of the points. Additionally, each processor sorts the cluster
centers according to their distances to a bounding box around the process-local points. Evaluating the target clusters in increasing distance order allows
to abort early when the minimum distance of the remaining clusters is above the distance of already found candidates.

\paragraph{Scalable Edge Partitioning} Schlag~\etal~\cite{DBLP:conf/alenex/Schlag0SS19} present a distributed algorithm to solve the edge partitioning problem.
The edge partitioning problem asks for a partition $\Pi = \{E_1, \ldots, E_k\}$ of the edge set into $k$ blocks each containing roughly the same number of edges, while minimizing the vertex cut
$\sum_{v \in V} \rho(v) - 1$ where $\rho(v) = |\{ E_i~|~E_i \in \Pi: I(v)~\cap~E_i \neq \emptyset \}|$.
They evaluated two methods to solve the problem. The first transforms the graph
into its dual hypergraph representation (edges of the graph become vertices of the hypergraph and each vertex of the graph induces a hyperedge spanning its incident edges). Using a hypergraph partitioner that optimizes the connectivity metric
to partition the vertex set directly optimizes the vertex cut of the underlying edge partitioning problem. The second method uses a distributed construction algorithm of the so-called \emph{split-and-connect} graph (SPAC).
For each vertex $u$, it inserts $d(u)$ auxiliary vertices into the SPAC graph and connects them to a cycle using auxiliary edges each with weight one. Each auxiliary vertex is a representative for exactly one
incident edge of $u$. For each edge $(u,v) \in E$, it adds an infinite weight edge between the two representatives of the corresponding edge. Thus, a partition of the vertex set of the SPAC graph cannot cut an
edge connecting two representatives.
Therefore, such a partition can be transformed into an edge partition
by assigning each edge to the block of its representatives.
In the evaluation, they compare both representations while using different graph and hypergraph partitioners.
The results showed that parallel graph partitioners outperform distributed hypergraph partitioners. However, the sequential
hypergraph partitioner KaHyPar~\cite{DBLP:phd/dnb/Schlag20} produces significantly better vertex cuts than all other approaches (more than $20\%$ better than the
best graph-based approach), but is an order of magnitude slower than the evaluated distributed algorithms.

\ifEnableExtendDONE
2013 JA-BE-JA: A Distributed Algorithm for Balanced Graph Partitioning \cite{DBLP:conf/saso/RahimianPGJH13} \\
\tobiasx{I have no idea whats going on here}
2015 A Scalable Distributed Graph Partitioner \cite{DBLP:journals/pvldb/MargoS15} \\
2017 (conf version 2015) Parallel Graph Partitioning for Complex Networks \cite{DBLP:journals/tpds/MeyerhenkeSS17}\\
2017 Social Hash Partitioner: A Scalable Distributed Hypergraph Partitioner \cite{socialhash} \\
2017 Partitioning Trillion-Edge Graphs in Minutes~\cite{xtrapulp} \\
2017 Spinner: Scalable Graph Partitioning in the Cloud \cite{DBLP:conf/icde/MartellaLLS17} \\
\lars{this might be interesting for cuts vs packing} \tobiasx{this paper does not present a distributed partitioner}
2017 Graph Partitioning for Distributed Graph Processing \cite{DBLP:journals/dase/OnizukaFS17} \\
2018 Balanced $k$-means for Parallel Geometric Partitioning \cite{DBLP:conf/icpp/LoozTM18} \\
2019 (conf 2016) Distributed Balanced Partitioning via Linear Embedding \cite{DBLP:journals/algorithms/AydinBM19} \\
2019 Scalable edge partitioning \cite{DBLP:conf/alenex/Schlag0SS19} \\
2020 Scalable, Multi-Constraint, Complex-Objective Graph Partitioning \cite{xtrapulp-journal} \\
2020 PMondriaan: A Parallel Hypergraph Partitioner \cite{berg2020pmondriaan}
\fi

\subsection{GPU}
Due to their high computational power, modern GPUs have become
an important tool for accelerating data-parallel applications.
However, due to the highly irregular structure of graphs, it remains
challenging to design graph algorithms that efficiently utilize the
SIMD architecture of modern~GPUs.

\paragraph{Multilevel Graph Partitioning}
Goodarzi~\etal~\cite{DBLP:conf/ipps/GoodarziBG16, DBLP:conf/ieeehpcs/GoodarziKSG19}
present two algorithms for GPU based multilevel graph partitioning.
Their earlier approach~\cite{DBLP:conf/ipps/GoodarziBG16} uses
heavy-edge matching for coarsening and transfers the coarsest graph
onto CPU for initial partitioning (using Mt-Metis~\cite{DBLP:conf/icpp/KarypisK95}).
During refinement, vertices are distributed among threads and each thread
finds the blocks maximizing the gain values of its assigned vertices.
To prevent conflicting moves that worsen the edge cut in combination,
refinement alternates between rounds in which only moves to blocks
with increasing (resp. decreasing) block IDs are considered.
For each block, potential moves to the block are collected in a global
buffer, which is then sorted, and the highest rated moves are executed.

Their later approach~\cite{DBLP:conf/ieeehpcs/GoodarziKSG19} brings
several improvements.
First, the authors use Warp Segmentation~\cite{DBLP:conf/IEEEpact/KhorasaniGB15}
to improve the efficiency of the heavy-matching computation during
coarsening.
Initial partitioning is then performed on the GPU using a  greedy growing
technique.
During refinement, vertices are once more divided among threads, and
each thread finds the blocks maximizing the gain of its assigned
boundary vertices and collects the potential moves in a global buffer.
Then, the algorithms finds the highest rated $\ell$ moves in the global buffer,
for some small input constant $\ell$.
Since moves might conflict with each other, their algorithm
distributes all $2^{\ell}$ move combinations across thread groups
and finds the best combination, which is then applied to the graph
partition.
This process is repeated until the global buffer is empty.
On average, their GPU based approach is approximately $1.9$ times faster
than Mt-Metis while computing slightly worse edge cuts across their
benchmark set of 16 graphs.

Fagginger Auer and Bisseling~\cite{Auer2012GraphCA} present a fine-grained shared-memory parallel algorithm for graph coarsening
and apply this algorithm in the context of graph clustering to obtain a fast greedy
heuristic for maximising modularity in weighted undirected graph. The algorithm is suitable for both multi-core CPUs and GPUs.
Later, Gilbert \etal \cite{9460473} present
performance-portable graph coarsening algorithms. In particular, the authors study a GPU parallelization of the heavy edge coarsening method. The authors evaluate their coarsening method using a multilevel spectral graph partitioning algorithm as primary use case.

\paragraph{Spectral Graph Partitioning}
The availability of efficient eigensolvers on
GPUs has lead to a recent re-emergance of spectral techniques for
graph partitioning on GPU systems~\cite{DBLP:conf/ipps/AcerBR20, sphynx-tr, nvr-2016-001-tr}.
These techniques were first developed by
Donath and Hoffman~\cite{donath1972algorithms, donath1973lower} and
Fiedler~\cite{fiedler1975property} to compute graph bisections in the
1970s.
Subsequently, these techniques have been
improved~\cite{Boppana87, pothen1990partitioning, simon1991partitioning, hendricksonSpectral95, BarSim93}
and extended to partition a graph into more than two blocks using
multiple eigenvectors~\cite{hendricksonSpectral95, DBLP:conf/dac/AlpertY95}.

Naumov~and~Moon~\cite{nvr-2016-001-tr} present an
implementation of spectral graph partitioning for single GPU systems
as part of the nvGRAPH library, whereas Acer~\etal~\cite{DBLP:conf/ipps/AcerBR20, sphynx-tr}
propose the multi-GPU implementation Sphynx.
Both partitioners precondition the matrix and use the LOBPCG~\cite{DBLP:journals/siamsc/Knyazev01}
eigenvalue solver.
The eigenvectors are then used to embed the graph into a multidimensional
coordinate space, which is then used to derive a partition of the graph.
In nvGRAPH, this is done using a $k$-means clustering algorithm on the
embedded graph, whereas Sphynx uses the geometric graph partitioner
Multi-Jagged~\cite{DBLP:journals/tpds/DeveciRDC16} which supports
multi-GPU systems.
Since the approach by Acer~\etal outperforms nvGRAPH in terms of
partition balance, cut size and running time (when run on a single
GPU system), we focus on their experimental evaluation.
To this end, when Sphynx is compared against ParMETIS~\cite{DBLP:journals/jpdc/KarypisK98},
ParMETIS generally obtains significantly better cuts than Sphynx ---
approx. 20\% (resp. 70\%) lower cuts on regular (resp. irregular) graph instances).
On irregular graphs, the authors report a significant speedup of
approximately 19 using Sphynx on 24 GPUs compared to ParMETIS with 168 MPI processes
across four compute nodes, although ParMETIS is approximately
3 times faster than Sphynx on regular graphs even when using a single
CPU core for each GPU used by Sphynx.
Additionally, Acer~\etal report the influence of several matrix
preconditioners on different classes of graphs.

\subsection{Approaches for Other Types of Hardware}
Recently, Ushijima-Mwesigwa~\cite{10.1145/3149526.3149531} \etal explore graph partitioning using quantum annealing on the D-Wave 2X machine. The main idea is to formulate the graph partitioning problem as a quadratic unconstraint binary optimization problem via a spectral approach. These problems can be mapped to minimizing Ising objective functions which the D-wave system can minimize. The method could run directly on the D-Wave system for small graphs and used a hybrid approach for large graphs.

Lio \cite{10.1007/978-3-031-08757-8_40} \etal experiment with solving the graph partitioning problem on the Fujitsu Digital Annealer which is a special-purpose hardware designed for solving combinatorial optimization problems. In particular, the authors also model the problem as a quadratic unconstraint binary optimization problem via a spectral approach.. The Fujitsu Digital Annealer then utilizes application-specific integrated circuit hardware for solving fully connected quadratic unconstraint binary optimization problems. The authors identify a dense network for which their approach significantly outperforms KaffpaE. Then a range of similar instances is created using the MUSKETEER~\cite{DBLP:conf/fusion/GutfraindSM15} framework. On most of those instances, their approach outperforms KaffpaE and Gurobi.

\section{Experimental Methodology}
\label{sec:experimentalmethodology}

As more applications of (hyper)graph partitioning have emerged in recent years, it becomes increasingly important to evaluate
algorithms on many (hyper)graph instances to demonstrate their effectiveness for practical applications. The presentation of results
using tables comparing the running time and solution quality of different algorithms can quickly become difficult to interpret, while evaluations
based on aggregated numbers can lead to misleading conclusions. This section discusses several alternative methods helpful
for presenting and interpretting experimental results.

\begin{figure*}[!t]
  \begin{minipage}{.99\textwidth}
  \ifpdfplots
    \includegraphics{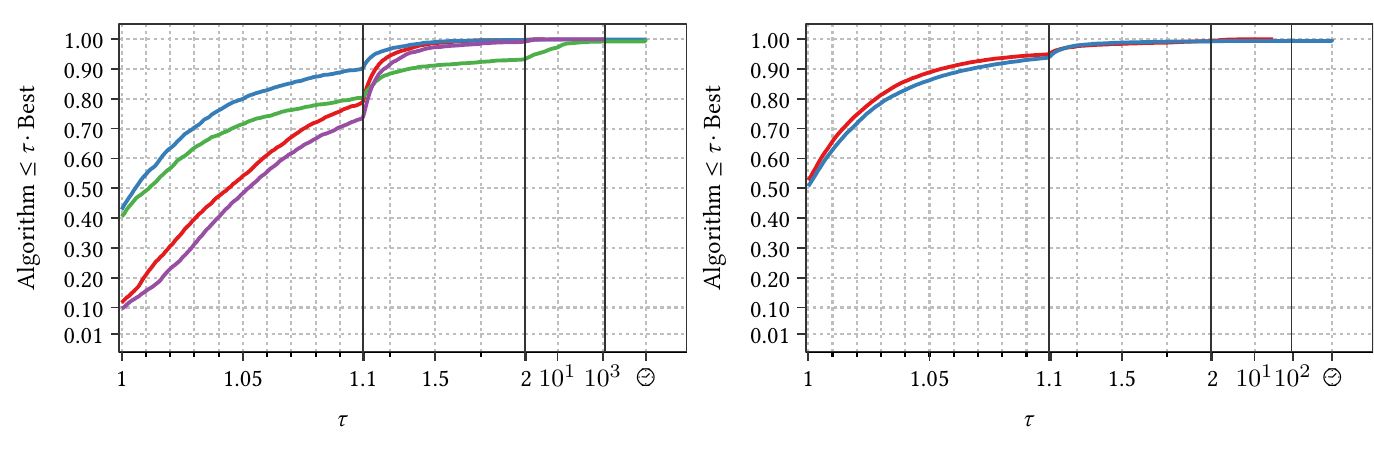}
  \else
    \tikzexternalenable
    \tikzsetnextfilename{img/performance_profile}%
    \input{img/performance_profile}%
    \tikzexternaldisable
  \fi
  \end{minipage} %
  \begin{minipage}{.99\textwidth}
    \vspace{-0.4cm}
    \centering
  \ifpdfplots
    \includegraphics{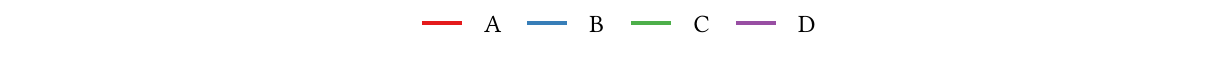}
  \else
    \tikzexternalenable
    \tikzsetnextfilename{img/performance_profile_legend}%
    \input{img/performance_profile_legend}%
    \tikzexternaldisable
  \fi
  \end{minipage} %
	\vspace{-0.4cm}
  \caption{Performance profiles comparing four different algorithms (left) and
           the result of a effectiveness test between two algorithms (right).
           The \ClockLogo~tick marks instances for which the corresponding algorithm
           has computed an infeasible solution or has ran into a time limit.}
  \label{fig:performance_profiles}
	\vspace{-0.5cm}
\end{figure*}

\paragraph{Performance Profile}
\emph{Performance profiles} can be used to compare the solution quality of different algorithms~\cite{DBLP:journals/mp/DolanM02}.
Let $\mathcal{A}$ be the set of all algorithms, $\mathcal{I}$ the set of instances, and $q_{A}(I)$ the value of a quality metric (e.g., edge-cut) of algorithm
$A \in \mathcal{A}$ on instance $I \in \mathcal{I}$.
For each algorithm $A$, performance profiles show the fraction of instances ($y$-axis) for which $q_A(I) \leq \tau \cdot \text{Best}(I)$, where $\tau$ is on the $x$-axis
and $\text{Best}(I) := \min_{A' \in \mathcal{A}}q_{A'}(I)$ is the best solution produced by an algorithm $A \in \mathcal{A}$ on an instance $I \in \mathcal{I}$.
For $\tau = 1$, the $y$-value indicates the percentage of instances for which an algorithm $A \in \mathcal{A}$ performs best.
Achieving higher fractions at smaller $\tau$ values is considered~better.

Figure~\ref{fig:performance_profiles} (left) compares a quality metric of four different algorithms using a performance profile.
Algorithm A and C compute on roughly $40\%$ of the instances the best solutions while algorithm B and D only on $10\%$ (see $\tau = 1$).
The solutions produced by algorithm A are worse than the best solutions by $\approx 4\%$ in the median (intersection of $y = 0.5$ with the red line is
at $\tau \approx 1.04$).
If we compare algorithm C and D based on their geometric mean of the
quality metric, we would observe that algorithm C is only $0.2\%$ better than D. Hence, we would probably conclude that there is no
significant difference between both.
If we look at the performance profile, we see that algorithm C is on most of the instances closer to the best solution than D.
However, on $\approx 10\%$ of the instances the solutions are worse than the best once
by more than a factor of two, which has large influence on its geometric mean (see green line at $\tau = 2$).

\paragraph{Effectiveness Tests}
The performance profile in Figure~\ref{fig:performance_profiles} (left) suggests that algorithm B produces solutions that are better than those of A.
However, if we look at their running times, we observe that algorithm A is more than an order of magnitude faster than B on average.
Thus, algorithm B may have an unfair advantage due to its longer running time.
Therefore, Akhremtsev \etal~\cite{mt-kahip-journal} introduces \emph{effectiveness tests} to compare solution quality when two algorithms are given a
similar running time, by performing additional repetitions with the faster algorithm.
Consider two algorithms A and B, and an instance $I$.
Effectiveness tests first sample one run of both algorithms.
Let $t_A^1$ and $t_B^1$ be their running times and assume that $t_A^1 \geq t_B^1$.
Then, additional runs of B are sampled until the accumulated time exceeds $t_A^1$.
Let $t_B^2, \dots, t_B^l$ denote their running times.
The last run is accepted with probability $(t_A^1 - \sum_{i = 1}^{l-1} t_B^i) / t_B^l$ so that the expected time for the runs of $B$ equals $t_A^1$.
The quality metric taken is the minimum out of all runs.

Figure~\ref{fig:performance_profiles} (right) shows the result of the effectiveness test for algorithm A and B using a performance profile.
The plot shows that there is no significant difference between the
two algorithms.

\paragraph{Significance Tests}
It is not immediately obvious from the performance profile in Figure~\ref{fig:performance_profiles} (left) whether algorithm B performs
better than C or vice versa since both lines are close to each other. Here, statistical tests can give further information
that helps to decide whether the difference between two algorithms is statistically significant.

A widely used technique is to formulate a \emph{null hypothesis} assuming that there is no difference between the
observed distributions. A significance test then calculates the probability ($p$-value) that the observed
distribution occurs under the assumption that the null hypothesis is true~\cite{young2005essentials}.
The null hypothesis is rejected if the probability is below a predefined significance level $\alpha$
(e.g, $\alpha = 5\%$ as suggested by Fisher~\cite{fisher1992statistical}).
To compute the $p$-value, the Wilcoxon signed rank test~\cite{WILCOXON} is often used to
compare two algorithms while the Friedman test~\cite{FRIEDMAN} makes pairwise comparisons between multiple algorithms.

Significance tests based on a null hypothesis decide if the difference between two measurements is
statistically significant but do not reveal any information whether or not the difference is relevant in practice.
For example, a significance test comparing the running times of two algorithms A and B may conclude that A is faster than B. However,
A might be only $1\%$ faster on average and, thus, the relevance of the improvement is questionable.
Therefore, Angriman~\etal~\cite{DBLP:journals/algorithms/AngrimanGLMNPT19} recommend using \emph{parameter estimation}
instead of hypothesis testing. To apply parameter estimation to our example, we could compute the running time ratios $t_A(I) / t_B(I)$ for each instance
$I$ and determine the \emph{confidence interval} that contains $95\%$ of the measurements. The confidence interval gives additional
information on the \emph{effect size} of the improvement.

\section{Future Challenges}
\label{sec:futchallenges}
While there has been a considerable progress in the field in the last decade, there is a wide range of challenges that remain open.
It is an interesting question to what extent the multitude of results sketched above have
reached a state of maturity where future improvements become less and less likely. Algorithms like multi-threaded graph partitioning for balanced graph partitioning using a small number of threads and standard inputs may be difficult to improve. Long time open problems like large gaps between theory and practice may remain open for a long time. On the other hand, other important issues have considerable potential. We try to identify some of them below.

\emph{Parallelism and Other Hardware Issues.} Scalability of high quality parallel (hyper)graph partitioning remains an active area of research. In particular, achieving good scalability and quality on large distributed memory machines is still a challenge, but even on shared-memory machines, scalability to a large number of threads seems difficult. Even more difficult is aligning the inherent complexity and irregularity of state-of-the-art algorithms with the restrictions of GPUs or SIMD-instructions. Another conundrum is that, for good memory access locality during partitioning, (hyper)graphs need to already be partitioned reasonably well.

Hierarchies of supercomputers have to be taken into account during partitioning. This can be done by using multi-recursive approaches taking the system hierarchy into account or by adapting the deep multilevel partitioning approach sketched above to the distributed memory case. When arriving at a compute-node level, additional techniques are necessary to employ the full capabilities of a parallel supercomputer. For example, many of those machines have GPU's on a node level. Recently, researchers started to develop partitioning algorithms that run on GPUs and while of independent interest, partitioning algorithms developed for this type of hardware can help in that regard. Hence, future parallel algorithms have to compute partitions on and for heterogeneous machines. On the other hand, algorithms should be energy-efficient and performance per watt has to be considered. Lastly, future hardware platforms have to be taken into consideration when developing such algorithms. One way to achieve this will be to use performance portable programming ecosystems like the Kokkos~library~\cite{kokkos}.

\emph{Problem Variations.} Current partitioning algorithms work well on a wide range of instances. Most of these instances are either unweighted and translate into well behaved weighted problems if a multilevel algorithm is used or instances have a fairly even distribution of node weights. For instances with vastly different values of node weights, partitioning problems discussed in this paper get close to bin packing problems. Currently solvers are not able to handle this case~well.

Algorithms for the well established objective functions like edge cut (in the graph case) or connectivity/cut (in the hypergraph case) are able to compute very high-quality partitions. Many open problems remain when optimizing for other or multiple objectives and when the problem has other/multiple constraints.
For example, in parallel computing, bottleneck objectives such as minimizing the maximum edge cut of a block should perform better, but few partitioners optimize for such objectives.
The same is true for high-quality solvers for problems with multiple constraintsin the case of repartitioning.
Streaming algorithms currently compute much worse quality partitions than internal memory partitioning algorithms. This is partially due to the fact that such algorithms don't have a global view of the optimization problem.
Improving quality of streaming algorithms is an open problem. Lastly, we believe that directed variants of the problems will become more important in order to model different applications.

\emph{Multilevel Partitioning.} The multilevel technique has been incredibly successful in the field of decomposition. Yet, multilevel algorithms in the area still consist in practice
of a very limited number of multilevel techniques.
Development and understanding components of more sophisticated coarsening schemes,
edge ratings, and metrics of nodes’ similarity that can be propagated throughout the hierarchies are among the future challenges for (hyper)graph partitioning as well as any attempt of their
rigorous analysis. For example, coarsening at the moment is typically stopped based on simple formulas. However, in practice it may be much more fruitful to derive stopping rules for coarsening that take the given instance into account. This also translates to different parameters of sub-algorithms that should be chosen based on the type of input that is provided to the algorithm.

\emph{Modeling and Experiments.} Traditionally instances for partitioning algorithms came from scientific simulations and thus were fairly well structured. For these networks, edge cut and communication volume of the application have been highly correlated. However, for instances having much more complex structures like social networks that recently became important in practice, the correlation is not that high. It remains an open question how different models of (hyper)graph partitioning translate into concrete
          performance of applications.

\textbf{Acknowledgements.}
We would like to thank Ilya Safro for feedback and additional references on an early version of the manuscript. 
We acknowledge support by DFG grants SCHU 2567/1-2, WA654/19-2, SA933/10-2, SA933/11-1,
and NSF grant CCF-1919021.

\vspace*{-.25cm}
\bibliographystyle{myplainurl}
\bibliography{references.bib,phdthesiscs.bib}

\end{document}